\documentclass[reprint,aps,pre,superscriptaddress]{revtex4-1}

\usepackage{amsmath,graphicx,multirow,array}%,todonotes,acro,subcaption}
\usepackage{amsthm,amssymb}
\usepackage{CJK}

\newcommand{\pfrac}[2]{\frac{\partial{#1}}{\partial{#2}}}

\usepackage[usenames,dvipsnames]{color}
\usepackage[normalem]{ulem}
\begin{document}
\begin{CJK*}{UTF8}{}

\title{Films, layers and droplets: The effect of near-wall fluid structure on spreading dynamics}

\author{Hanyu Yin (\CJKfamily{gbsn}尹寒玉)}
\affiliation{Department of Mathematical Science,
Loughborough University, Loughborough, LE11 3TU, UK}

\author{David N. Sibley}
\affiliation{Department of Mathematical Science,
Loughborough University, Loughborough, LE11 3TU, UK}

\author{Uwe Thiele}
\affiliation{Institut f\"ur Theoretische Physik, Westf\"alische Wilhelms-Universit\"at M\"unster, Wilhelm Klemm Str.\ 9, 48149 M\"unster, Germany}
\affiliation{Center of Nonlinear Science (CeNoS), Westf{\"a}lische Wilhelms-Universit\"at M\"unster, Corrensstr.\ 2, 48149 M\"unster, Germany}
\affiliation{Center for Multiscale Theory and Computation (CMTC), Westf{\"a}lische Wilhelms-Universit\"at, Corrensstr.\ 40, 48149 M\"unster, Germany}

\author{Andrew J. Archer}
\affiliation{Department of Mathematical Science,
Loughborough University, Loughborough, LE11 3TU, UK}

\date{\today}

\begin{abstract}
We present a study of the spreading of liquid droplets on a solid substrate at very small scales. We focus on the regime where effective wetting energy (binding potential) and surface tension effects significantly influence steady and spreading droplets. In particular, we focus on strong packing and layering effects in the liquid near the substrate due to underlying density oscillations in the fluid caused by attractive substrate-liquid interactions. We show that such phenomena can be described by a thin-film (or long-wave or lubrication) model including an oscillatory Derjaguin (or disjoining/conjoining) pressure, and explore the effects it has on steady droplet shapes and the spreading dynamics of droplets on both, an adsorption (or precursor) layer and completely dry substrates. At the molecular scale, commonly used two-term binding potentials with a single preferred minimum controlling the adsorption layer height are inadequate to capture the rich behaviour caused by the near-wall layered molecular packing. The adsorption layer is often sub-monolayer in thickness, i.e., the dynamics along the layer consists of single-particle hopping, leading to a diffusive dynamics, rather than the collective hydrodynamic motion implicit in standard thin-film models. We therefore modify the model in such a way that for thicker films the standard hydrodynamic theory is realised, but for very thin layers a diffusion equation is recovered.
\end{abstract}

\maketitle

\end{CJK*}

\section{Introduction}
The spreading of liquid droplets is a fascinating and highly consequential phenomenon which has received great attention for over a century \cite{de1985wetting, sui2014numerical, snoeijer2013moving, bonn2009wetting}. When a small volume of liquid is placed on a solid substrate, it can spread to form a hemispherical drop with a free surface and three-phase equilibrium contact angle $\theta$. This is referred to as partial wetting. However, if the liquid molecules are strongly attracted to the substrate the liquid spreads as much as it can, forming a pancake shaped ultrathin drop. This is referred to as complete wetting. This wetting behaviour influences phenomena that arise in everyday life, such as in the sliding of rain drops on windows or plant leaves, paint coating a wall or tear films in the eye \cite{de2013capillarity}. As well as being a simple day to day process, static and dynamic wetting behaviour also influences many industrial processes. Critical applications such as coating, printing and lubrication have motivated many scientists to understand the evolution of thin liquid films and drops on substrates and to develop models for their dynamics \cite{Duss1979arfm,bonn2009wetting}. All these wetting phenomena are governed by surface and interfacial interactions that occur over length scales varying from the very small (\r{A}) molecular distances to a few nm for van der Waals or electrostatic forces to the mesoscopic ($\mu$m) scale for capillary forces. Understanding the interplay of all these interactions and their influence on the interfacial fluid dynamics and thermodynamics is at the core of understanding the behaviour and properties of droplets and thin liquid films on a solid substrate.

Here, we develop a thin-film model in the form of a partial differential equation (PDE) that describes the time evolution of the local amount of liquid on a substrate. It includes many aspects of the microscopic interactions between the liquid molecules and the substrate. Note, that from the microscopic (statistical mechanics) viewpoint, \textit{adsorption} is a better defined and arguably more useful measure of the amount of liquid on the substrate than \textit{film thickness} \cite{hughes2015liquid}. This is because when the amount of liquid on the substrate is small (sub monolayer), then talking about a film height that is a fraction of a molecule does not make physical sense, whilst the adsorption is well-defined. In fact, when the vapour pressure is non-zero, the adsorption on the substrate can in principle even be negative, so in this case, talking about a film height is meaningless.

 Consider a fluid confined in a volume ${\cal V}$, in contact with a planar substrate with area ${\cal A}$ and with density distribution $\rho(\textbf{r})$. The total adsorption on the substrate is calculated from the fluid density profile as 
\begin{equation}
\hat{\Gamma} = \frac{1}{\cal A} \int_{\cal V}^{}  (\rho(\textbf{r})-\rho_b) \ \textrm{d}\textbf{r},
\end{equation}
where $\rho_b$ is the bulk fluid density, which in our context is the vapour density $\rho_v$.

If the planar substrate is assumed to correspond to the $z=0$ plane in a Cartesian coordinate system, then we can define the local adsorption as
\begin{equation}
\Gamma(x,y) = \int_{0}^{\infty} [\rho(x,y,z)-\rho_v] \ \textrm{d}z,
\end{equation}
which is roughly proportional to the film height $h$, since $\Gamma\approx h(\rho_l-\rho_v)$. In fact, following Ref. \cite{hughes2015liquid}, one can define the film height
\begin{equation}
h(x,y)\equiv \frac{\Gamma(x,y)}{(\rho_l-\rho_v)},
\label{eq:h_def}
\end{equation}
where $\rho_l$ is the density of the bulk liquid.

\begin{figure}[t]
\centering
\includegraphics[width=\columnwidth]{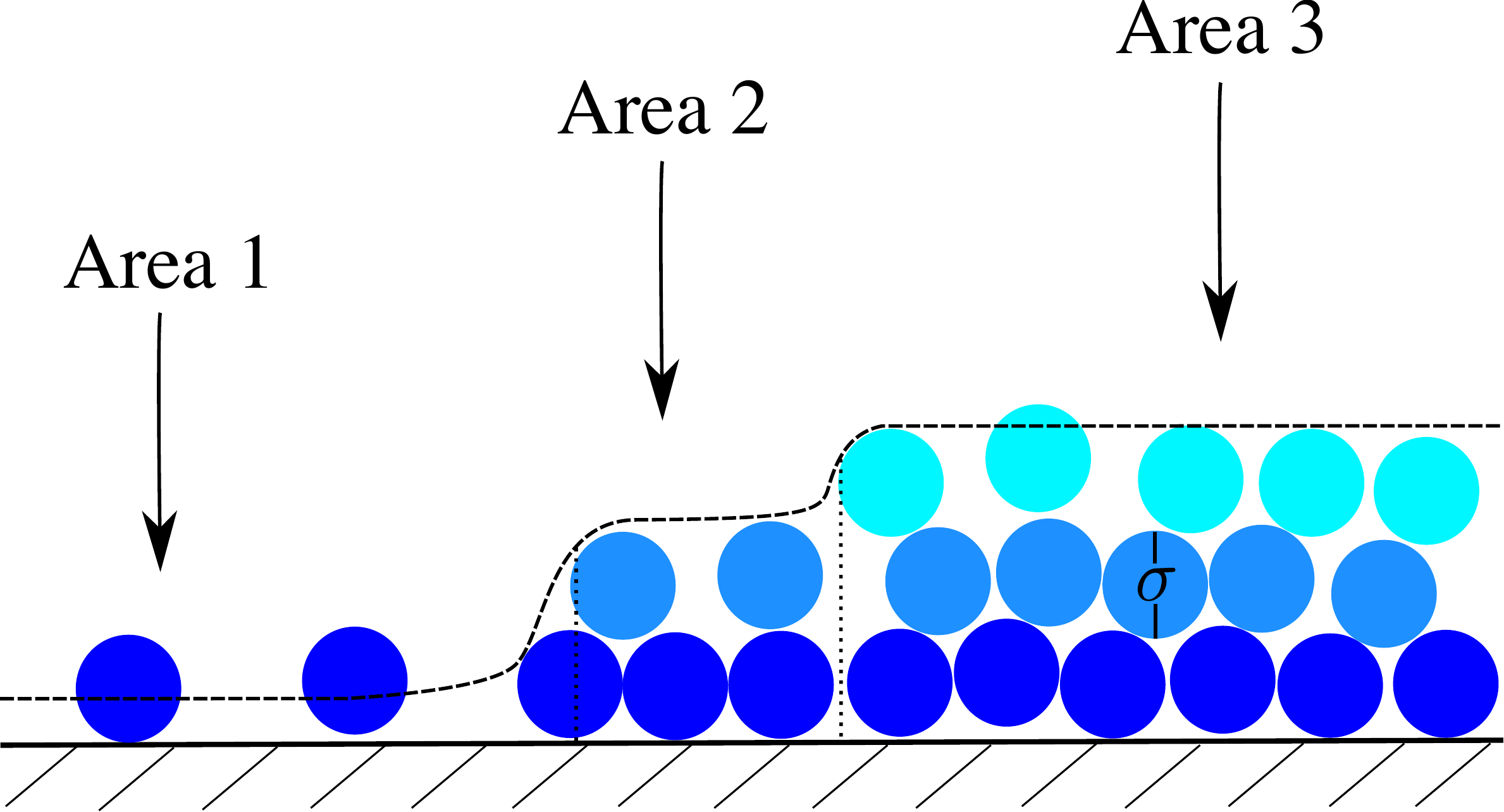}\caption{\label{adsorption} A sketch of the molecular configurations in a cross section through the contact line region of a drop of liquid which exhibits strong layering effects. The contact angle is $\theta\approx0$. We identify three distinct regions, Areas 1-3, where the adsorption takes three distinct values due to the fact that when the substrate adsorption is low, the influence of molecular packing becomes important. The dashed line gives the corresponding effective film height. In Area 1, the amount adsorbed on the substrate is low, so the effective film height $h\ll\sigma$, the diameter of the molecules.}
\end{figure}

When the cohesive forces between the molecules that form the liquid are short ranged compared to the size of the molecules $\sigma$, then strong layering at the substrate is possible, particularly at low temperatures \cite{de2013capillarity, brader2003statistical, heslot1989molecular, YaKB1992pra, DeCh1974jcis,KrDe1995cpl, hughes2015determining, hughes2016liquid}. See also the molecular dynamics simulation results in Ref.~\cite{isele2016requirements}. Consider the contact line region of a drop of such a system, that is close to the wetting transition, with $\theta\approx 0^\circ$. A sketch of a configuration of the molecules is displayed in Fig.~\ref{adsorption}. Three distinct regions, Areas 1--3 can be identified, based on the amount adsorbed. The thickness of the individual layers is approximately equal to the diameter of the liquid molecules, $\sigma$. In Area 1, there are just a few molecules adsorbed on the substrate, so the film height $h\ll\sigma$, as defined via Eq.\ \eqref{eq:h_def}. In Areas 2 and 3, the film height is roughly an integer number of molecular layers, since the strong intermolecular attractions favour complete layers. In Refs.\ \cite{hughes2015determining, hughes2016liquid}, density functional theory (DFT) for a simple model system was used to calculate the density distribution of a liquid at a substrate exhibiting this type of layered structure formation, which is remarkably similar to the terraced spreading drops observed in the experiments reported in \cite{heslot1989molecular}. In Refs.\ \cite{hughes2015determining, hughes2016liquid}, using the method developed in Ref.\ \cite{hughes2015liquid} the binding (or wetting) potential $g(h)$ was also calculated. This binding potential $g(h)$, together with the interfacial tensions, gives the excess free energy for having a liquid film of thickness $h$ adsorbed on the substrate (see Eq.~\eqref{g} below). It was found that in the types of situations sketched in Fig.\ \ref{adsorption}, the binding potential is oscillatory. In this paper we consider the influence of such an oscillatory binding potential on the shape of steady drops and also on the dynamics of drop spreading. The established thin film models describe the advective motion {\footnote{Note that here by `advective' motion, we refer to classical hydrodynamic motion of the film---principally using the term to contrast with the diffusive dynamics added in Sec.~\ref{sec:IV}}} of the liquid over the substrate, sometimes also incorporating slip \cite{bonn2009wetting,MWW2005jem}. However, normally, such models do not include the diffusive particle-hopping dynamics that one should expect when the adsorption is low \cite{ala2002collective,BOCM1996pre}, such as in Area~1 in Fig.~\ref{adsorption}. Thus, we also develop here an augmented thin-film equation that incorporates this effect, {with the principal aims of this work being (a) incorporating well-founded structural disjoining pressures into thin-film modelling, and (b) to propose and probe a model that switches between diffusion and hydrodynamics.}

This paper is structured as follows: The relevant physical concepts of interfacial science are introduced in Sec.~\ref{sec:wett}. In Sec.~\ref{sec:III} the mathematical description of steady and spreading drops is  derived and the solution methodology that is used to solve the model are introduced. An extension to include diffusive effects into the dynamics is discussed in Sec.~\ref{sec:IV} and results are presented in Sec.~\ref{sec:results}. Finally, our concluding remarks are made in Sec.~\ref{sec:VI}.

\section{Wetting behaviour and the form of the binding potential}
\label{sec:wett}

When a liquid drop is placed onto a dry solid substrate, it spreads a certain extent until equilibrium is reached and the free energy of the system is minimised. The extent of the spreading is determined by the equilibrium contact angle $\theta$, given by Young's equation \cite{young1805essay}
\begin{equation}\label{y}
\gamma_{lv} \cos \theta = \gamma_{sv} - \gamma_{sl},
\end{equation}
where $\gamma_{sv}$, $\gamma_{sl}$ and $\gamma_{lv}$ are the interfacial tensions (excess free energy per unit area) of the solid/vapour, solid/liquid, and liquid/vapour interfaces, respectively. These are defined when three phases are in equilibrium with each other \footnote{Note that as discussed in Ref.~\cite{de1985wetting}, $\gamma_{sv}$ is distinct from $\gamma_{so}$, the interfacial energy of the completely dry interface.}. 
For large enough droplets, $\theta$ is the inner angle the liquid-vapour interface makes with the substrate.

Complete wetting occurs when $\theta = 0^{\circ}$. The system is in equilibrium when a uniform macroscopically thick liquid layer covers the whole solid substrate. Partial wetting occurs when $ 0^{\circ}<\theta<180^{\circ}$. In this case, small droplets form a spherical cap due to dominant capillary effects. Deviations from a spherical cap shape occur if the radius of the drop is larger than the capillary length $\kappa = \sqrt{\gamma_{lv}/{(\rho g)}}$, so that gravity can no longer be neglected. Unless the vapour pressure of the liquid is zero, at equilibrium the substrate surrounding the drop is covered by a microscopically thin layer of thickness $h_0$ adsorbed on the substrate. This adsorbed layer is generally sub-monolayer [c.f.~Fig.~\ref{adsorption}], so the popular terminology `precursor film' is potentially misleading. For non-wetting to occur, $\theta = 180^{\circ}$. The wetting behaviour is determined by the molecular interactions (e.g.\ van der Waals, electrostatic forces, etc.). For very small drops and for length scales below a few hundred nm these interactions become important and can extend across the thickness of the film and contribute extra forces that determine the shape of the drop and how it spreads. For a film of thickness $h$, these interactions result in additional contributions to the free energy $g(h)$, often referred to as the binding potential. It can be expressed in terms of the Derjaguin (or disjoining/conjoining) pressure \cite{derjaguin1936anomalien, derjaguin1978question},
\begin{equation}
\Pi (h)= - \frac{\partial g(h)}{\partial h}.
\end{equation}
The interaction between the two interfaces can also be discussed in terms of this pressure. The total excess free energy per unit area of a system with a film of liquid with uniform thickness $h$ is:
\begin{equation}\label{g}
\frac{\bigtriangleup F}{\cal A} = \gamma_{lv} + \gamma_{sl} + g(h),
\end{equation}
where ${\cal A}$ is the area covered by the film. Thus, $g(h)$ gives the contribution to the free energy from the interactions between two interfaces, and has the limiting values $g(\infty) = 0$ and \cite{de2013capillarity}
\begin{equation}\label{new}
g(h_0) =  \gamma_{sv} - (\gamma_{lv} + \gamma_{sl}).
\end{equation}
The disjoining pressure gives the difference between the pressure $P$ in a thin liquid layer with thickness $h$ and $P(\infty)$, the pressure when the liquid film is macroscopically thick. Thus, 
\begin{equation}
\Pi(h) = P(h) - P(\infty).
\end{equation}

In general, three main contributions to the disjoining pressure $\Pi$ are identified \cite{Chen1984jcis,israelachvili1992entropic,davis1996statistical,Israelachvili2011}: (i) a long-range van der Waals contribution due to the interaction between dipoles (either permanent or induced) of the liquid molecules and the substrate molecules. It can be attractive or repulsive; (ii) Long-range electrostatic forces, from charge double layers overlapping during thinning. The like charged surfaces of the film repel each other;
 (iii) Short-range steric forces, which stem from the repulsive interactions between molecules when they are pushed close together.

The van der Waals force is characterised by the Hamaker constant $H$ \cite{israelachvili1992entropic,davis1996statistical,Israelachvili2011}, originating from the attractive (London) potential between individual pairs of molecules which at large $r$ is $\sim r^{-6}$, where $r$ is the distance between molecules. It gives the longest range contribution to $\Pi$. The leading order term that dominates for large $h$ is
\begin{equation}\label{eq:g_VdW}
\Pi \approx -\frac{H}{6\pi h^3}.
\end{equation}
If $H > 0$, it means the two interfaces attract each other and the film thins; on the other hand if $H < 0$, the two interfaces repel each other. Thus, the van der Waals force often determines the wetting behaviour. In the case $H>0$, one must include additional terms in $\Pi$. A commonly used expression that describes a partially wetting situation and allows for a stable precursor film is
\begin{equation}
\Pi_1(h) = \frac{5a}{h^6} - \frac{2b}{h^3},
\end{equation}
where $b=H/{12\pi}$ and $a$ is a positive constant. The corresponding binding potential is
\begin{equation}\label{eq:Pi}
g_1(h) = \frac{a}{h^5} - \frac{b}{h^2}.
\end{equation} 
The positive term represents the short range repulsive forces, while the second term describes the longer range van der Waals forces contribution. This form has been frequently used in thin-film models, e.g., in Refs.~\cite{schwartz1998simulation,Pismenmeso,thiele2010thin}. However, as can be deduced from considering the $h\to0$ limit, it is clear that this expression is really only valid for large film thickness.

Similarly, for model systems with only short range forces one finds that for large $h$ the binding potential decays exponentially: $g(h)\sim a_1 e^{-h/\xi}+a_2 e^{-2h/\xi}+\cdots$, where $\xi$ is the bulk correlation length in the liquid phase wetting the wall and coefficients $a_i$ depend on the temperature \cite{dietrich88,schick90}. The progress made in Refs.\ \cite{hughes2015liquid, hughes2015determining, hughes2016liquid} was to develop a DFT based method for calculating $g(h)$, or strictly speaking $g(\Gamma)$, that is valid over the whole range of values of $h$. Note that one can also use a molecular dynamics computer simulation based method for calculating $g(h)$ \cite{md2006jcp,MacDowell2011,tretyakov2013parameter,MacDowell2014adcolintsci,md2015pre}. For simple Lennard-Jones like fluids it was shown \cite{hughes2015liquid, hughes2015determining, hughes2016liquid} that the following form gives a good fit to the binding potential over the whole range:
\begin{equation}\label{eq:adam_fit}
g(h)=\frac{H(e^{-p(h)}-1)}{12\pi h^2},
\end{equation}
where $p(h)=h^2(a_0e^{-a_1h}+a_2+a_3h+a_4h^2+a_5h^5)$. Eq.\ \eqref{eq:adam_fit} gives the correct decay for $h\to\infty$, namely that in Eq.\ \eqref{eq:g_VdW}, but remains finite in the limit $h\to0$. However, as also shown in Ref.\ \cite{hughes2015determining}, Eq.\ \eqref{eq:adam_fit} is not appropriate for all liquids as it does not capture any layering effects. It is then shown that for a simple fluid with only short-range attractive interactions between the molecules (i.e., no van der Waals contribution, $H=0$), the following form gives a good fit to the binding potential data obtained using DFT:
\begin{align}\label{bp2}
g_2(h) &= e^{-\frac{h}{a_0}}[a_5+a_4\cos(a_1h+a_2)] +a_6e^{-2\frac{h}{a_0}} \nonumber
\\
+& a_7e^{-3\frac{h}{a_0}}
+a_8e^{-4\frac{h}{a_0}}+a_9e^{-5\frac{h}{a_0}}+a_{10}e^{-6\frac{h}{a_0}}.
\end{align}
$a_0$ is the bulk correlation length in the liquid phase at the interface, and the other $a_i$'s corresponds to further constants, determined via fits to the DFT data and for the particular treated case take the following values: $a_0 = 0.907508$, $a_1 = -7.35183$, $a_2 = 5.90059$, $a_3 = 0$, $a_4 = -0.011038$, $a_5= -0.000147646$, $a_6 = 0.0449827$, $a_7 = 0.422683$, $a_8 = -0.7673$, $a_9 = -0.230683$ and $a_{10}= 0.559131$ \cite{hughes2015determining}. These are all in units where the particle diameter $\sigma=1$ and the thermal energy $k_BT=1$, where $k_B$ is the Boltzmann constant and $T$ the temperature of the system. Henceforth, these are the units in which all lengths and energies are given.

\begin{figure}[t]
\centering
\includegraphics{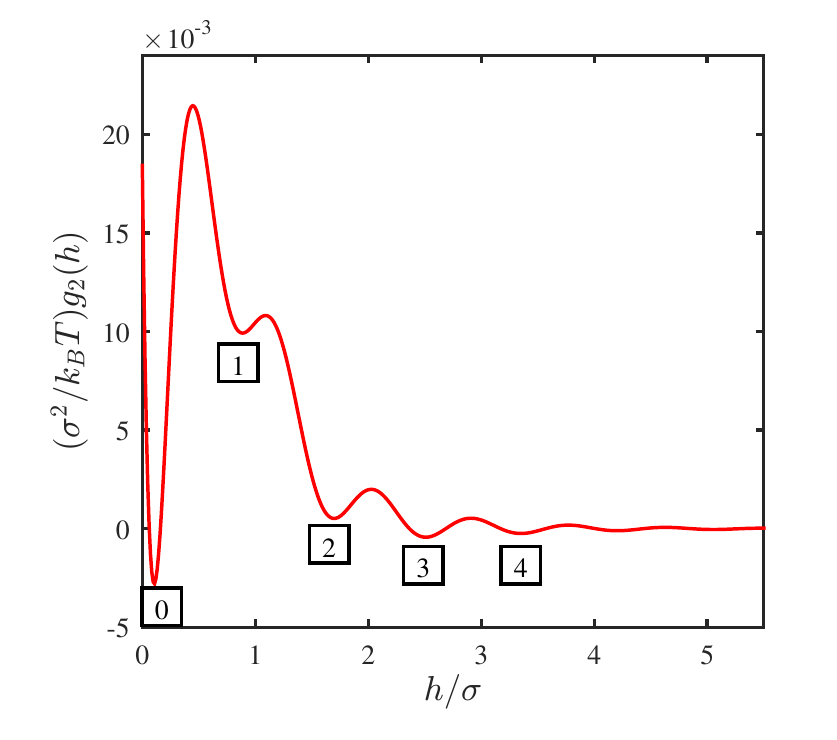}
\caption{\label{bp2figure} A plot of the binding potential in Eq.\ (\ref{bp2}), calculated in \cite{hughes2015determining} using DFT. The global minimum is labelled `$0$', as this state corresponds to Area 1 in Fig.~\ref{adsorption}, where the film thickness is almost zero. The local minima at larger $h$ are labelled `1, 2, 3, 4' and represent one, two, three and four layers of molecules, respectively.}
\end{figure}

Eq.~\eqref{bp2} has a damped oscillatory decay as $h\to\infty$. The oscillations lead to the presence of multiple minima in $g_2(h)$, which result in the formation of `steps' or `terraces' in the vicinity of the contact line at the droplet edge -- examples are displayed later in Sec.~\ref{sec:results}. Each subsequent minimum in $g_2(h)$ corresponds to the addition of a further layer. In Fig.\ \ref{bp2figure} we display a plot of $g_2(h)$, appropriately scaled with $\sigma^2/(k_BT)$. The global minimum is labelled `0', where the film thickness (adsorption) is very small, and there are almost zero molecules on the substrate, so the system is partially-wetting, but with small contact angle $\theta$, since $g(h_0)$ is only slightly negative. The boxed labels `1', `2', `3', and `4' indicate the local minima corresponding to the respective number of complete layers of molecules on the substrate. We can see from Fig.~\ref{bp2figure} that in this system, one layer of molecules is not as favourable as two or more complete layers of molecules. Note that the oscillatory behaviour in $g_2(h)$ is also seen in the corresponding liquid density profiles in the full DFT calculations \cite{hughes2015determining}.

\begin{figure}[t]
\centering
\includegraphics{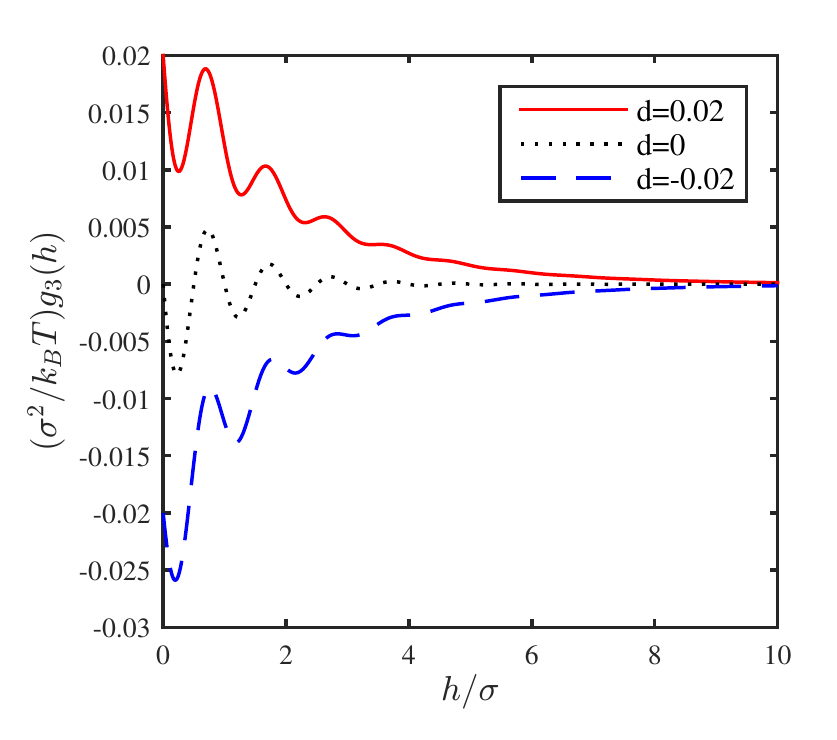}
\caption{\label{bp3} A plot of binding potential $g_3(h)$ in Eq.\ (\ref{bp3equation}) with $a = 0.01$,
  $b=\pi/2$,
  $c=1$,
  $k=2\pi$,
  $d=0.02$ (solid red line) $d=0$ (dotted black line)  and $d=-0.02$ (dashed blue line). }
\end{figure}

\begin{figure}[t]
\centering
\includegraphics{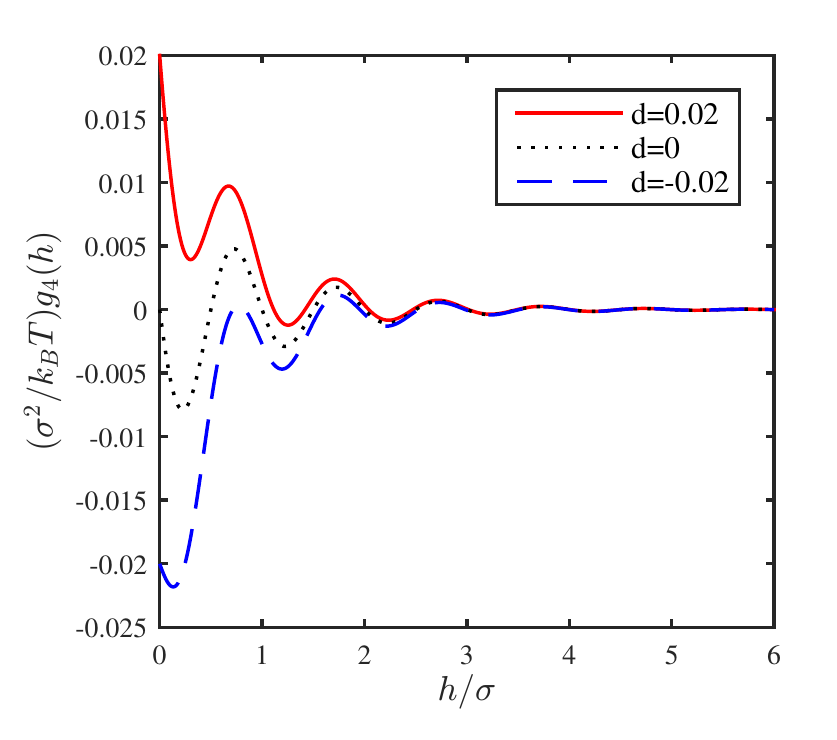}
\caption{\label{bp4} A plot of binding potential $g_4(h)$ in Eq.\ (\ref{g4}) with $a = 0.01$,
  $b=\pi/2$,
  $c=1$,
  $k=2\pi$,
  $d=0.02$ (solid red line) $d=0$ (dotted black line)  and $d=-0.02$ (dashed blue line). }
\end{figure}

Eq.\ (\ref{bp2}) contains many parameters and is the binding potential for a particular liquid on a particular substrate at a particular temperature \cite{hughes2015determining}. Here we seek to understand the overall effects of oscillatory binding potentials on liquid drop shapes and the spreading behaviour. Therefore we truncate the expression in Eq.~\eqref{bp2} to obtain the following simplified expressions,
\begin{equation}\label{bp3equation}
g_3 (h) = a\cos (hk + b)e^{-\frac{h}{c}} + d e^{-\frac{h}{2c}},
\end{equation}
and
\begin{equation}\label{g4}
g_4 (h) = a\cos (hk + b)e^{-\frac{h}{c}} + d e^{-\frac{2h}{c}},
\end{equation}
where $a$, $b$, $c$, $d$, $k$ are coefficients that we vary to determine the generic types of behaviours that one can observe. These simpler expressions retain the overall character of the expression in Eq.~\eqref{bp2}, but contain fewer parameters. In Figs.~\ref{bp3} and \ref{bp4} we display plots of the binding potentials (\ref{bp3equation}) and (\ref{g4}), respectively, with the typical parameter values that we use in our study, namely $a = 0.01$, $b=\pi/2$, $c=1$, $k=2\pi$, and varying the parameter $d$, such as $d=0.02$ (red) which represents a wetting situation, $d=0$ (black) refers to a partially-wetting case but close to the wetting transition, and $d=-0.02$ (blue) partially-wetting.

These two binding potentials $g_3(h)$ and $g_4(h)$ are somewhat more generic than $g_2(h)$, but at the same time they retain the oscillatory behaviour of $g_2(h)$, which gives the layering. The lowest (positive) local minimum of $g_3(h)$ for $d=0.02$ is at $h = 0.2522$ which is similar to the local minimum of the full expression $g_2(h)$ with the parameter values obtained from the DFT results in Ref.\ \cite{hughes2015determining}, namely $h = 0.1081$. For relatively large $h$, both $g_3(h)$ and $g_4(h)$ tend to zero but with different limiting behaviours. For $g_3(h)$, the exponential decay dominates whereas for $g_4(h)$, the sinusoidal oscillations dominate. Note that the ultimate asymptotic decay is determined by the form of the decay into bulk of the liquid density profiles that are in contact with a substrate \cite{evans1993asymptotic, evans1994asymptotic, hughes2015determining, hughes2016liquid}. Whether the decay is monotonic or damped oscillatory depends on the state point (i.e.\ temperature and density) and on which side of the Fisher-Widom (FW) line this state point is. The FW line is the locus in the phase diagram at which the asymptotic decay of the radial distribution function crosses over from monotonic to damped-oscillatory decay \cite{evans1993asymptotic, evans1994asymptotic, fisher1969decay}.

\section{Thin film equation}
\label{sec:III}

The time evolution of a thin liquid film on a surface can be derived from energetic considerations. The free energy of the system is
\begin{equation}
\label{eq:ourF}
F[h] = \iint \left[ g\left(h\right) + \frac {\gamma_{lv}}{2} \left(\nabla h\right)^2\right] \mathrm{d}x \mathrm{d}y.
\end{equation}
This contains two contributions: (i) the binding potential contribution from the molecular interactions with the substrate and (ii) the energy of the free surface (surface tension), where $\nabla = (\frac{\partial}{\partial x} , \frac{\partial }{\partial y})$ is the 2D gradient operator. The latter is proportional to the fluid surface area and the approximation $\sqrt{1+(\nabla h)^2}\approx  1+\frac{1}{2}(\nabla h)^2$, appropriate when the gradients are small, has been made in Eq.\ \eqref{eq:ourF}. We have also neglected an irrelevant constant term.

The quantity
\begin{equation}\label{eq:dF_dh}
\frac{\delta F}{\delta h} = -\Pi - \gamma_{lv}\nabla^2h
\end{equation}
is the negative of the local pressure in the film and so any gradients in this quantity give the thermodynamic force which drives the flow of liquid over the substrate. There is therefore a current $\textbf{j} = -Q(h)\nabla \frac{\delta F}{\delta h}$, where $Q(h)$ is the mobility coefficient. Combining this with the continuity equation, we obtain \cite{mitlin1993dewetting,thiele2010thin}
\begin{equation}\label{diffusioneq}
\frac{\partial h}{\partial t} = - \nabla\cdot \textbf{j} = \nabla . \left[Q(h)\nabla \frac {\delta F[h]}{\delta h}\right].
\end{equation}
The mobility coefficient $Q(h)$ depends on the film thickness. Often  the expression $Q(h)= h^3/(3\eta)$ is assumed, where $\eta$ is the fluid viscosity. This is what emerges from the long-wave approximation of the Navier-Stokes equations with no-slip boundary conditions \cite{oron1997long}, giving
\begin{equation}\label{final}
\frac{\partial h}{\partial t} =  \nabla \cdot\left(\frac{h^{3}}{3\eta} \nabla\left(-\gamma_{lv}\nabla^2 h - \Pi\left(h\right)\right)\right).
\end{equation}
Assuming some slip, then $Q(h)$ can acquire additional terms, for example, Navier-slip \cite{navier1823memoire} results in $Q(h) = \beta h^2 + h^3/3\eta$ \cite{oron1997long, munch2005lubrication}. In the next section we discuss further mobilities $Q(h)$ which describe diffusion effects.

\begin{figure}[t]
\includegraphics{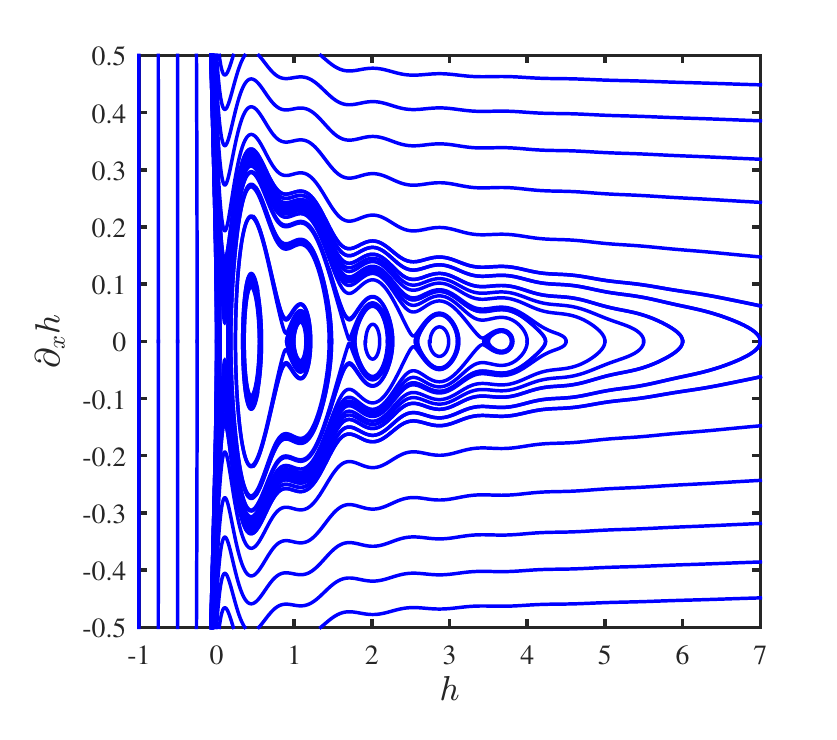}
\caption{\label{stream} A phase plane diagram when the binding potential is $g_2$ in Eq.~\eqref{bp2} and $\gamma_{lv} = 0.51$, which shows all possible equilibrium solutions of our thin film equation as the maximal drop height is varied, with the Lagrange multiplier having the value $\lambda=9.8 \times 10^{-4}$.}
\end{figure}

Equilibrium (steady state) drop profiles are those which minimise $F[h]$ subject to the constraint that the volume of the liquid $V=\iint h  \mathrm{d}x \mathrm{d}y$ is fixed, i.e.\ which minimise
\begin{equation}
\Omega[h]\equiv F [h]+ \lambda \iint h  \mathrm{d}x \mathrm{d}y,
\end{equation}
where $\lambda$ is the Lagrange multiplier associated with the volume constraint. The minimising curve satisfies
\begin{equation}
\frac{\delta \Omega}{\delta h} = 0,
\end{equation}
which is equivalent to
\begin{equation}\label{ga}
-\Pi - \gamma_{lv} \nabla^2 h = \lambda.
\end{equation}
From this we can identify $\lambda$ as the pressure difference across the interface due to the Laplace and disjoining pressures. If we consider a 1D droplet such that $h=h(x)$ and let $u=h, v=u'$, we have
\begin{align}
u'&=h',\\
v'&=\frac{1}{\gamma_{lv}}\left(\frac{dg}{dh}-\lambda\right),
\end{align}
which can be used to plot the phase plane diagram of equilibrium solutions, as shown in Fig.~\ref{stream} (for standard Derjaguin pressures such plots can be found in Refs.~\cite{mitlin1993dewetting,TVNP2001pre} where also the influence of $\lambda$ is discussed). In this figure closed loops correspond to periodic solutions -- i.e. the solutions that one obtains on finite domains with periodic boundary conditions. This allows one to determine at a glance the range of equilibrium profiles that one may expect to obtain from the model. The oscillations in the `streamlines' in Fig.~\ref{stream} correspond to steps or terraces in the contact line region of the droplet solutions. Such droplet profiles are obtained below as stationary solutions of our PDE model.

\section{Incorporating the effects of diffusion into the mobility}
\label{sec:IV}

In droplet spreading simulations with the advective mobility, we see that on some occasions a very thin precursor film can extend ahead of the bulk of the main droplet. Physically, when this film is very thin, in particular when the film thickness is much less than the order of a particle diameter $\sigma$, such as illustrated in Area 1 in Fig.~\ref{adsorption}, we would expect the motion to be dominated by diffusion. Currently our model only describes the advective motion over the substrate -- recall that Eq.~($\ref{final}$) is obtained from a long-wave approximation to Navier-Stokes plus no-slip boundary conditions. 

We now seek to modify the thin-film equation in a simple way such that both, diffusion and advection, occur throughout the droplet, but for where the adsorbed film is very thin (i.e.\ less than roughly a monolayer or two), diffusion dominates, but for where the film thickness is larger, advection dominates. Thus, the two limiting cases we require are: $(i)$ the diffusion equation
\begin{equation}\label{diffe}
\pfrac{h}{t} = \nabla \cdot (D \nabla h ), \qquad \mbox{when } h\ll \sigma,
\end{equation}
where $D$ is the diffusion coefficient for particles moving over the surface, and $(ii)$ the thin-film equation
\begin{equation}\label{thinfilme}
\pfrac{h}{t} = \nabla \cdot \left[\frac{h^3}{3\eta} \nabla \frac{\delta F}{\delta h}\right], \qquad \mbox{when } h\gg \sigma,
\end{equation}
which is Eq.~$(\ref{final})$. Both Eqs.~(\ref{diffe}) and (\ref{thinfilme}) can be obtained as appropriate limits of the more general Eq.~$(\ref{diffusioneq})$, if the mobility $Q(h)$ is suitably generalised, as we now show.

To see how a diffusion equation can be obtained, we consider the case when $h$ is close to $0$. From the Maclaurin series expansion of $g(h)$ we obtain $-\Pi(h) = g'(0)+g''(0)h + \ldots$, which, together with Eq.~\eqref{eq:dF_dh}, gives
\begin{equation}\label{eq:DF_dh_series}
\frac{\delta F}{\delta h} = g'(0)+ g''(0)h - \gamma_{lv}\nabla^2h+\ldots.
\end{equation}
If we assume
\begin{equation}\label{eq:Q_general}
Q(h) = \alpha_0 + \alpha_1 h + \alpha_2 h^2 + h^3/(3\eta),
\end{equation}
then in the $h\to0$ limit we have $Q(h) \approx \alpha_0$ and so from Eqs.~\eqref{diffusioneq} and \eqref{eq:DF_dh_series} we obtain
\begin{equation}
\frac{\partial h}{\partial t} \approx  \alpha_0 \nabla^2 \left(g''(0)h-\gamma_{lv}\nabla^2 h \right),
\end{equation}
which in the limit where the first term dominates (for example if scaling into regions near contact lines), yields Eq.~\eqref{diffe}, together with the result that $\alpha_0=D/g''(0)$.

In the limit when $h$ is large, Eq.~\eqref{eq:Q_general} gives $Q(h) \approx h^3/(3\eta)$, and so the desired result, Eq.~\eqref{thinfilme}, is recovered. Thus, our final model is
\begin{align}\label{everything}
\frac{\partial h}{\partial t} = & \nabla \cdot\left[ \left( \frac{D}{g''(0)}
+\frac{h^{3}}{3\eta}\right) \nabla\left(-\gamma_{lv}\nabla^2 h - \Pi\left(h\right)\right)\right],
\end{align}
where we have set the coefficients $\alpha_1$ and $\alpha_2$ to zero. If we kept the terms involving $\alpha_1$ and $\alpha_2$ we would effectively be also investigating the effect of slip at the substrate, which at these molecular scales is usually seen as a coarse grained method to account for any number of physical processes that allow for contact line motion \cite{my_Crack}. In particular, as mentioned previously $\alpha_2$ can be associated with the popular Navier-slip model, and $\alpha_1$ with a nonlinear slip model \cite{munch2005lubrication}. Detailed comparisons between slip models in the thin-film setting can be found elsewhere \cite{HaleyMiksis, SavvaPrecursorSlip, My_JEM}.

We note that other authors have discussed the modelling of diffusion or diffusive regimes in  thin films \cite{yochelis2005droplet,popescu2012precursor,honisch2015instabilities}. Ref.~\cite{yochelis2005droplet} used a piece-wise mobility for diffusive and advective regimes in a mesoscopic hydrodynamic approach to droplet motion due to surface freezing/melting. The authors of Ref.~\cite{popescu2012precursor} have a discussion on adiabatic and diffusive films, but they talk about them in a different context than the present work. In \cite{popescu2012precursor} the argument is that when the edge of the film is very thin it becomes approximately flat so that curvature effects are negligible. In this case the thin film equation can be written in the same form as the diffusion equation with an (approximately constant) height-dependent diffusion term being $D(h) = -{h^3}/(3\eta) \partial_h\Pi$. Clearly this is quite different to our proposed implementation where we wish to model a diffusive region of the droplets where height increases from zero or a negligible value $h\ll\sigma$ to $h\approx\sigma$ rapidly. Finally, Ref.~\cite{honisch2015instabilities} compares the time evolution of relaxing liquid ridges employing various different mobility functions, including a diffusive one.

Returning to our governing equation \eqref{everything}, we now nondimensionalise, by scaling
\begin{gather}
\nabla=\frac{1}{\sigma}\nabla^*, \quad h=\sigma h^*, \quad t=\tau t^*, \quad \Pi=\frac{k_BT}{\sigma^3}\Pi^*, \nonumber \\ \gamma_{lv}=\frac{k_BT}{\sigma^2}\gamma_{lv}^*, \quad F =  {k_BT}F^*,
\label{eq:nondimscals}
\end{gather}
where we recall that $\sigma$ is the diameter of particles on the substrate, and $\nabla^*$, $h^*$, $t^*$, $\Pi^*$ and $\gamma_{lv}^*$ are the dimensionless quantities, and we have also given a scaling for the total free energy as its dynamic evolution is investigated in our numerical results presented below. By taking
\begin{gather}
\tau = \frac{3\eta\sigma^3}{k_BT}, \quad \bar \alpha_0=\frac{3\alpha_0 \eta}{\sigma^3} = \frac{3 D \eta}{g''(0) \sigma^3},
\end{gather}
we obtain
%\begin{widetext}
\begin{equation}\label{eq:ge_use}
\frac{\partial h^*}{\partial t^*}=\nabla^*\cdot\left[\left(\bar\alpha_0
%+\bar\alpha_1h^*+\bar\alpha_2(h^*)^2
+h^{*3}\right)\nabla^*\left(-\gamma_{lv}^*\nabla^{*2} h^*-\Pi^*\right)\right].
\end{equation}
%\end{widetext}
Subsequently, we drop the `*' for simplicity. This is our new equation which describes the evolution of the liquid film, incorporating both advection and diffusion.

\section{Results and discussion}
\label{sec:results}

\subsection{Spreading on a non-zero background adsorption}

The results presented in this section are obtained for the four binding potentials introduced in section \ref{sec:wett}. The time simulations are initiated at the instant when the droplet is released onto the substrate and are carried out until the droplet comes to rest, at the equilibrium steady state. The initial condition is assumed to be a drop shape which is modelled with a Gaussian function of the form
\begin{equation}\label{initialcondition}
h(x, t = 0) = C e^{-\left[\frac{(x-x_f/2)}{E}\right]^2} + h_b,
\end{equation}
where the parameter $h_b$ is the background value of $h$, i.e.\ the imposed layer height far away from the droplet, which can be set to a value $\neq h_0$ if desired. Recall that the height $h_0$ is the height of the `precursor' film/foot that extends away from the droplet during its approach to equilibrium. It is the height at which the lowest (positive) minimum of the binding potential occurs. In particular, $h_0$ corresponds to the global minimum of $g_1$ (the commonly investigated spreading situation) and $g_2$ (the terraced spreading situation), but not necessarily the global minimum of $g_3$ or $g_4$ (the generalised versions of $g_2$), depending on their parameter values.

In Eq.~\eqref{initialcondition}, the parameter $C$ is the amplitude of the initial droplet, i.e.\ it is the height of the droplet above the background film at time $t=0$. $E$ controls the width of the initial droplet, and $x_f$ is the length of the domain.
 
This initial condition specifies the height $h(x, t=0)$ in the $z$-axis, and it is assumed uniform in the $y$-direction to create a 2D droplet. The spreading of 2D droplets has been investigated extensively \cite{oron1997long,My_JEM} as they give qualitatively the same behaviour as for 3D axisymmetric droplets, particularly in the important region near the contact line. Periodic boundary conditions are applied, so that $h(x=0,t)=h(x=x_f,t)$, and the symmetry of the droplet is preserved in the dynamics due to the symmetric initial condition of Eq.~(\ref{initialcondition}).

All computations are performed using a method of lines technique, using finite difference approximations for the spatial derivatives, trapezoidal numerical integration for computing integrals (for the free energy and for confirming mass conservation), and the {\it ode15s} Matlab variable-step, variable-order (VSVO) solver \cite{matlabodesuite}. {A convergence test was applied, with the conclusion that a small enough grid size should be applied---typically here $dx=0.2$,} details are given in the Appendix.

To be able to compare the spreading dynamics of droplets with the two binding potentials $g_1$ and $g_2$, given by Eqs.~(\ref{eq:Pi}) and (\ref{bp2}) respectively, we first must find the values of parameters $a$ and $b$ in $g_1$ which make the contact angle and $h_0$ for both binding potentials the same. %\ttcoaut{Text deleted from here}.
Combining Eq.~(\ref{new}) and Young's equation (\ref{y}), gives the following relationship between the minimum of the binding potential and the equilibrium contact angle \cite{de1985wetting,rauscher2008wetting, churaev1995contact}
\begin{equation}\label{calc}
\theta=\cos^{-1}\left(1+\frac{g(h_0)}{\gamma_{lv}}\right).
\end{equation}

For $g_2$ in Eq.~\eqref{bp2}, with the coefficients given by DFT calculation \cite{hughes2015determining}, the minimum of $g_2$ is $-0.0028$ at $h_0=0.1081$, and the corresponding surface tension (also from the DFT \cite{hughes2015determining}) is $\gamma_{lv}=0.5101$. This value of $\gamma_{lv}$ is kept fixed throughout this section to enable a fair comparison between other effects such as the form of binding potential. The effect of surface tension is to smooth out gradients in the liquid-vapour film height $h$, thus a smaller surface tension would enhance the influence of the oscillatory binding potentials and give sharper terraces at equilibrium. Substituting these values back to Eq.~(\ref{calc}) gives the equilibrium contact angle $\theta=6.006^\circ$. By equating the value and location of the minimum of $g_1$ in terms of the parameters $a$ and $b$, with the respective numeric values for $g_2$, we find $a=2.756 \times 10^{-8}$ and $b=5.453\times 10^{-5}$, which allow a direct comparison between binding potentials $g_1$ and $g_2$, and the investigation of the effect of oscillatory binding potentials.

\begin{figure}[t]
\centering
\includegraphics{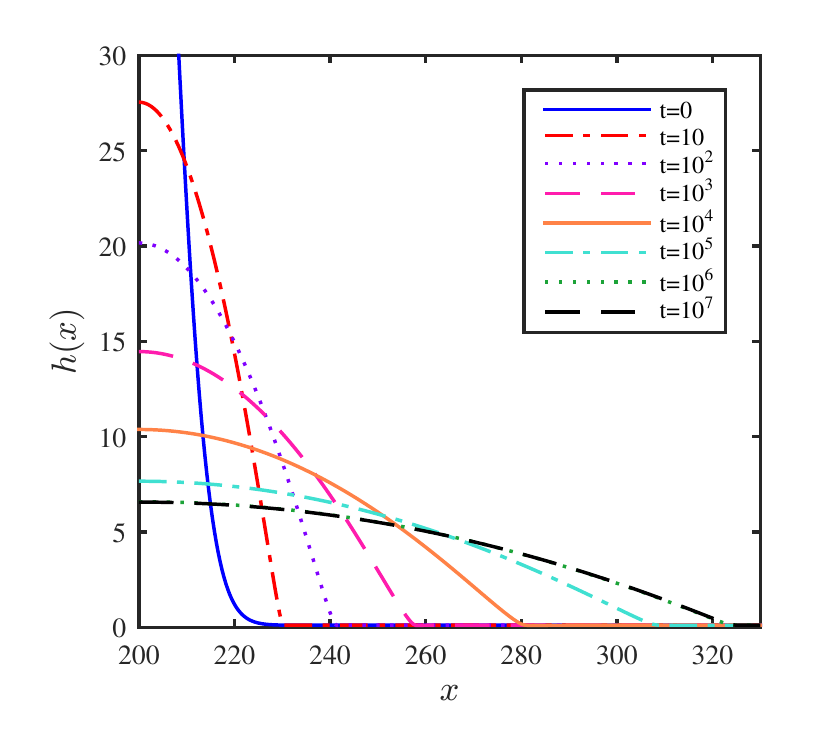}
\caption{\label{bigpic2} A time sequence of drop profiles for a liquid drop spreading on a substrate, with binding potential $g_1$ given by Eq.~(\ref{eq:Pi}) with $a=2.756\times 10^{-8}$ and $b=5.453\times 10^{-5}$. The parameters in the initial condition are chosen as $h_b = 0.1081$, $C = 60$ and $E = 10$. The $t=0$ profile is centred at $x=200$ and then spreads until reaches equilibrium. The diffusive mobility coefficient $\bar\alpha_0=0$.}
\end{figure}

Fig.~\ref{bigpic2} shows the time sequence of drop profiles for a liquid drop spreading on a substrate already covered by an equilibrium background film, with binding potential $g_1$ given by Eq.~(\ref{eq:Pi}), with $a=2.756 \times 10^{-8}$ and $b=5.453\times 10^{-5}$. The diffusive mobility coefficient $\bar\alpha_0=0$. The parameters in the initial condition are chosen as $h_b = 0.1081$, $C = 60$ and $E = 10$. The $t=0$ profile is centred at $x=200$ and then spreads. The liquid spreads rapidly from rest until $t\approx10^4$ under the effect of surface tension due to the significant difference between effective imposed initial contact angle and the equilibrium contact angle. The spreading then slows as the droplet equilibrates, as expected. The curves for times $t = 10^6$ and $t = 10^7$ are overlapped and virtually indistinguishable, indicating that equilibrium has been reached.

\begin{figure}[t]
\centering
\includegraphics{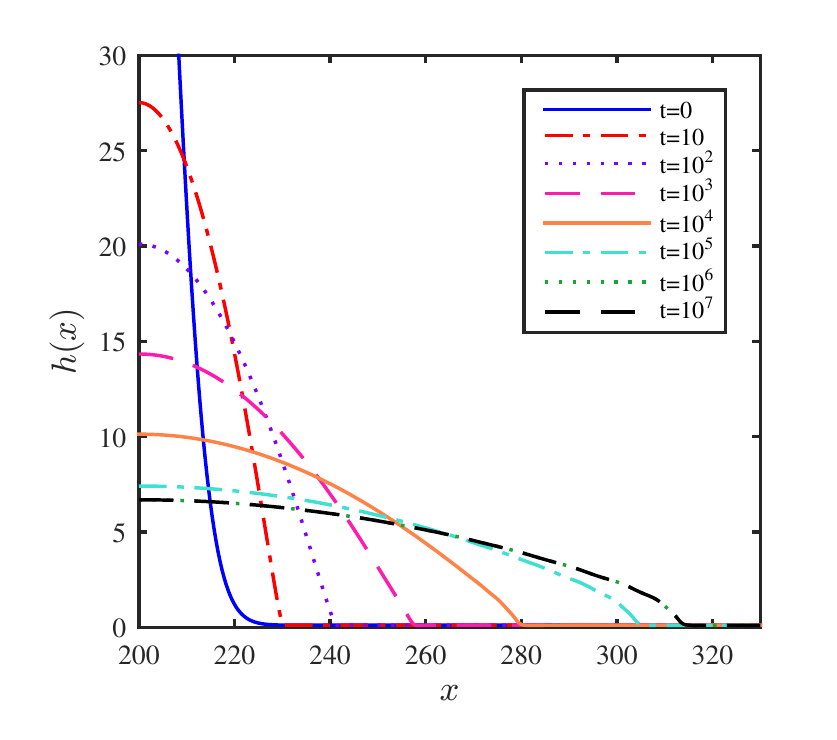}
\caption{\label{something1} A time sequence of drop profiles for a liquid drop spreading on a substrate, with binding potential $g_2$ given by Eq.~$(\ref{bp2})$. The initial condition is the same as for Fig.~\ref{bigpic2} and also $\bar\alpha_0=0$. The $t=0$ profile is centred at $x=200$ and then spreads.}
\end{figure}

The results in Fig.~\ref{bigpic2} are for the single-well binding potential of $g_1$. Fig.~\ref{something1} shows the equivalent time sequence of drop profiles for the oscillatory binding potential $g_2$ in Eq.~$(\ref{bp2})$. The spreading dynamics initially follows a very similar trajectory until approximately $t\approx10^5$. The height at the centre of the droplet continues to equilibrate at a similar rate to the case in Fig.~\ref{bigpic2} (with $g_1$) but for $t\gtrsim10^4$ we notice the emergence of a foot, or terrace, in the droplet near the contact line at a value $h\approx 2$ corresponding to the third minimum in the oscillatory binding potential $g_2$ (with label `2' in Fig.~\ref{bp2figure}), corresponding to a thickness of two particle layers. 

\begin{figure}[t]
\centering
\includegraphics{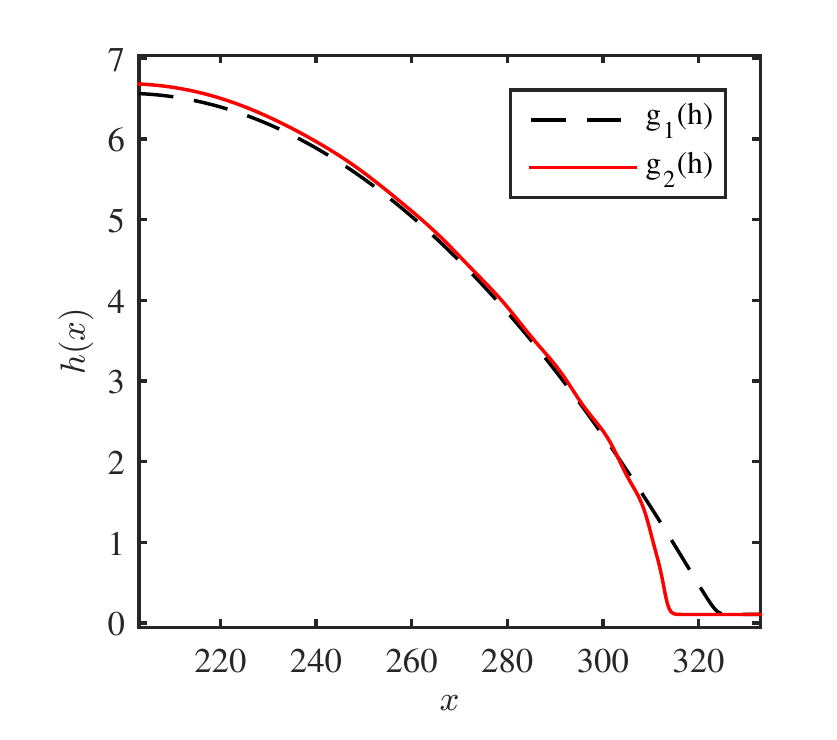}
\caption{\label{comparefinal} A comparison of the final equilibrium droplets with binding potentials $g_1$ and $g_2$, corresponding to Figs.~\ref{bigpic2} and \ref{something1}. The equilibrium contact angle $\theta$, the surface tension $\gamma_{lv}$ and the initial condition are the same for both cases.}
\end{figure}

From these comparisons we see that the dynamics is predominately driven by the relaxation of the contact angle to its equilibrium, and finer details of the binding potential do not dramatically change the timescales of spreading---especially when monitoring the equilibration of the maximal droplet height. However, significant differences near the contact line can occur where oscillations lead to terracing of the droplet. We see a reduction in wetting length for equilibrium droplets: the right contact line location at approximately $x=325$ and $x=315$ in Figs.~\ref{bigpic2} and \ref{something1}, respectively. A more detailed comparison of the two equilibrium droplet profiles is given in Fig.~\ref{comparefinal}. We see a marginal difference in maximal height, with more pronounced differences in the contact line region, with the final drop shape with binding potential $g_2$ having obvious `steps', or terraces. This is caused by the oscillations in the binding potential $g_2$, with each `step' corresponding to one layer of fluid particles (recall that whilst one layer is a local minimum in $g_2$, it is far less preferable than for $h=h_0$ or for two or more complete layers of particles).

\begin{figure}[t]
\centering
\includegraphics{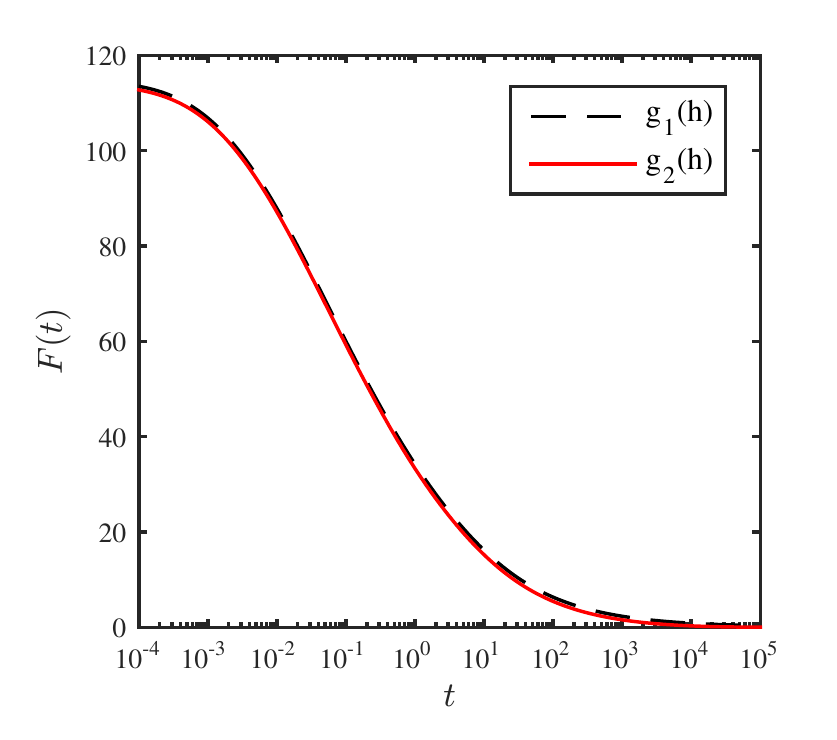}
\caption{\label{loglog2}The time evolution of the free energy \eqref{eq:ourF} as a droplet spreads to equilibrium with binding potentials $g_1$ and $g_2$, corresponding to Figs.~\ref{bigpic2} and \ref{something1}. The equilibrium contact angle $\theta$, the surface tension $\gamma_{lv}$ and the initial condition are the same for both cases. }
\end{figure}

Fig.~\ref{loglog2} shows the time evolution of the total free energies given by Eq.~(\ref{eq:ourF}), appropriately nondimensionalised as in Eq.~(\ref{eq:nondimscals}), corresponding to the two spreading situations in Figs.~\ref{bigpic2}--\ref{something1}. This shows that the timescales of spreading are unaffected by the choice of $g_1$ or $g_2$ in the case where coefficients were chosen to fix identical $h_0$ values and equilibrium contact angles. In this particular example then, it appears as though the formation of the terraces seen in Fig.~\ref{comparefinal} (with $g_2$) does not change the speed of approach to equilibrium. However, further investigation with a variety of initial conditions has shown that other events unique to oscillatory binding potentials can have significant effects, which we detail later.

Having demonstrated that oscillatory binding potentials are worthy of greater scrutiny through visualising the difference between the commonly used $g_1$ form to the specific $g_2$ one taken from a particular DFT calculation, we now analyse the more generic behaviour of binding potentials with oscillations, as given by the simplified forms $g_3$ and $g_4$.

\begin{figure}[t]
\centering
\includegraphics{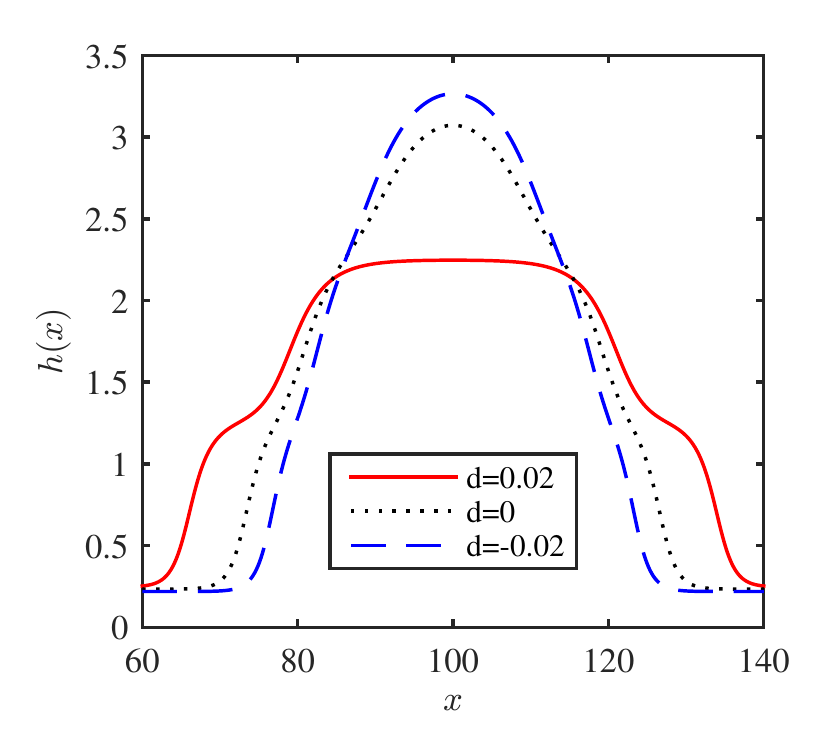}
\caption{\label{compareg3} Equilibrium drop profiles with binding potential $g_3$ in Eq.~(\ref{bp3equation}) as the parameter $d$ is varied and with $a = 0.01$, $b = \frac{\pi}{2}$, $c = 1$, $k = 2 \pi$.}
\end{figure}

In Fig.~\ref{compareg3} we display equilibrium drop profiles when the binding potential is $g_3$ in Eq.~\eqref{bp3equation}, for various values of the parameter $d$, namely $d= \{0.02,0,-0.02\}$ (c.f.\ Fig.~\ref{bp3}). The other parameters take the values $a = 0.01$, $b = \frac{\pi}{2}$, $c = 1$ and $k = 2  \pi$. The parameters in the initial condition are chosen as $C=6$, $E=10$, $x_f=200$, and $h_b=\{0.2522, 0.2282, 0.1921\}$ for $d=\{0.02, 0, -0.02\}$, respectively. These give the lowest (positive) local minimum at a similar value to that of $g_2$, to allow for direct comparison. As discussed in Sec.~\ref{sec:wett}, $d = 0.02$ is wetting whereas the other two are partially-wetting (although $d=0$ is close to the wetting transition). From Fig.~\ref{compareg3} we see the influence of this, as the drops for higher $d$ spread out further, and thus have lower maximal height. The height spacing of the steps seen in these equilibrium drop profiles also corresponds to the spacing of the minima in $g_3$.

\begin{figure}[t]
\centering
\includegraphics{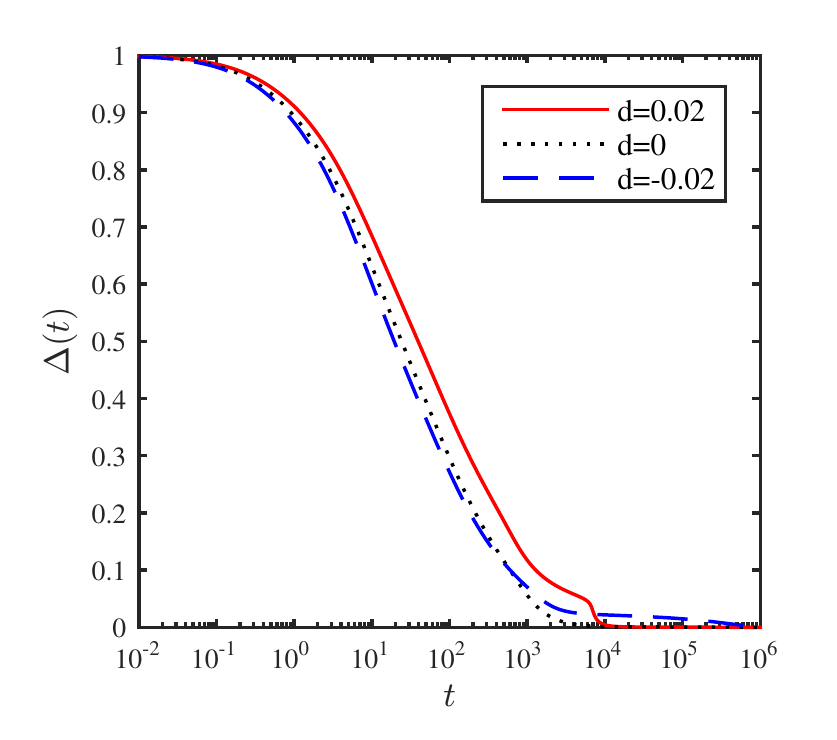}
\caption{\label{g3dne} The normalised free energy difference for droplet spreading under the influence of binding potential $g_3$, for various values of the parameter $d$. The expression of $g_3$ is given by Eq.~(\ref{bp3equation}), with $a = 0.01$, $b = \frac{\pi}{2}$, $c = 1$, and $k = 2 \pi$.}
\end{figure}

\begin{figure}[t]
\centering
\includegraphics{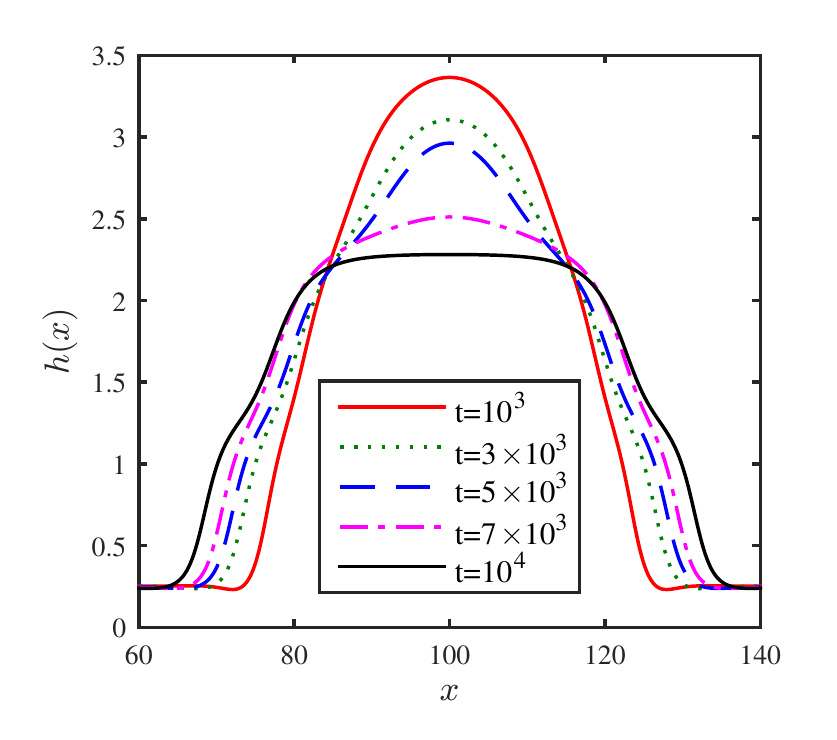}
\caption{\label{timeg3} A time sequence of drop profiles for a liquid drop spreading on a substrate, with binding potential $g_3$ given by Eq.~(\ref{bp3equation}), with $a = 0.01$, $b = \frac{\pi}{2}$, $c = 1$, $k = 2 \pi$, $d = 0.02$. The initial condition is chosen as $h_b=0.2522$, $C=6$, $E=10$ and $x_f=200$. The corresponding free energy time evolution is displayed as the solid (red) line in Fig.~\ref{g3dne}.}
\end{figure}

To explore the dynamics of spreading using $g_3$, in Fig.~\ref{g3dne} we plot the time evolution of the normalised free energy difference
\begin{equation}
\Delta(t)=\frac{F(t)-F(t=\infty)}{F(t=0)-F(t=\infty)},
\end{equation}
where the free energy $F$ is given by Eq.~(\ref{eq:ourF}). One might expect a smooth approach to equilibrium, as observed in Fig.~\ref{loglog2}. Instead, we see a number of stages to the dynamics, including the usual initial spreading/relaxation from the initial condition; the formation of terraces; a `popping' event where a rapid reduction of the free energy results from a sudden jump of the droplet free surface from one minimum of the binding potential to another; and then finally the usual long-time approach to equilibrium. The small jump for $d=0.02$ happens at $t\approx10^4$ and there are not any obvious jumps for $d=0$ and $d=-0.02$. 

To understand the dynamics more clearly, in Fig.~\ref{timeg3} we plot the time sequence of drop profiles for binding potential $g_3$ with $d=0.02$. The time is chosen from where the `popping' event is about to happen till it finishes. The drop starts to form two `terraces' from $t=10^3$, in the corresponding normalised free energy difference curve displayed in Fig.~\ref{g3dne} this behaviour is shown as the first inflection point where the normalised free energy difference $\approx 0.1$. As the drop spreads from $t=3\times10^3$ to $t=5\times10^3$, the top part of the drop becomes `sharper' and then it suddenly jumps down to form a flat top to minimise the free energy. This is an example of a `popping' events that leads to the sudden decreases observed in Fig.~\ref{g3dne}. After $t=10^4$ the drop keeps spreading and reaches equilibrium.

\begin{figure*}[t]
\centering
\includegraphics[width=0.68\columnwidth]{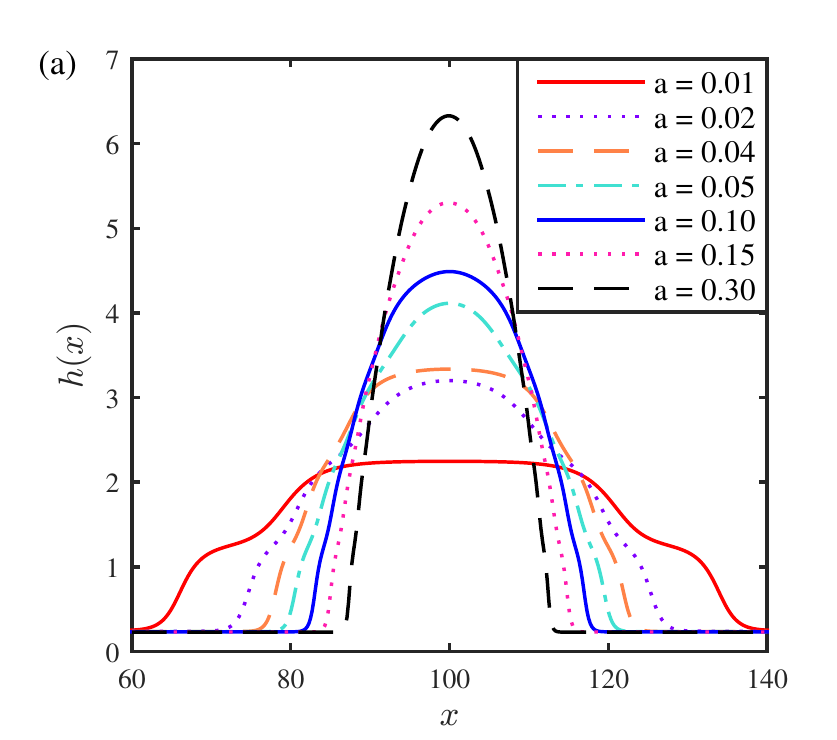}\includegraphics[width=0.68\columnwidth]{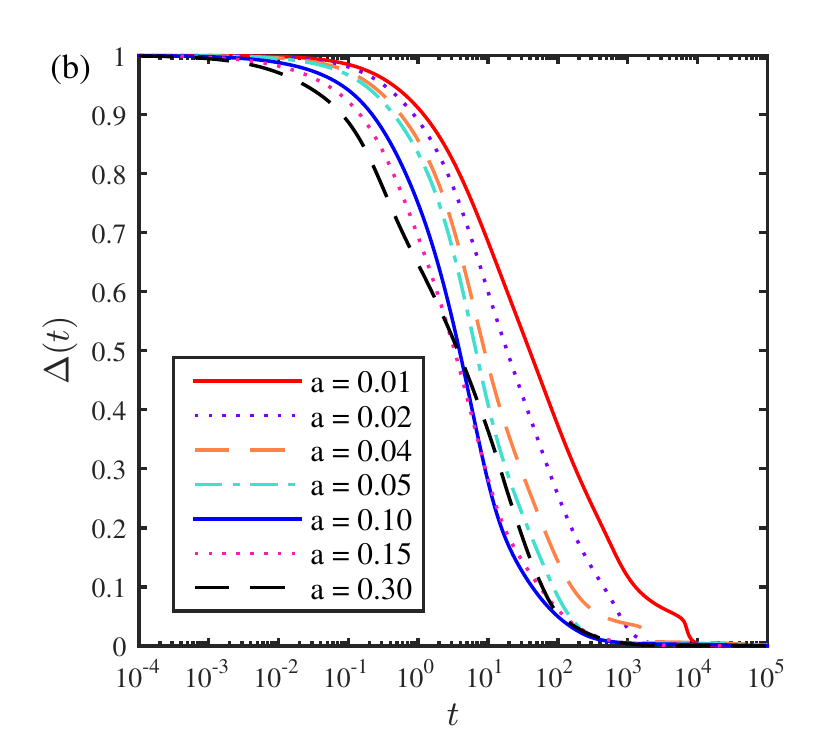}
\includegraphics[width=0.68\columnwidth]{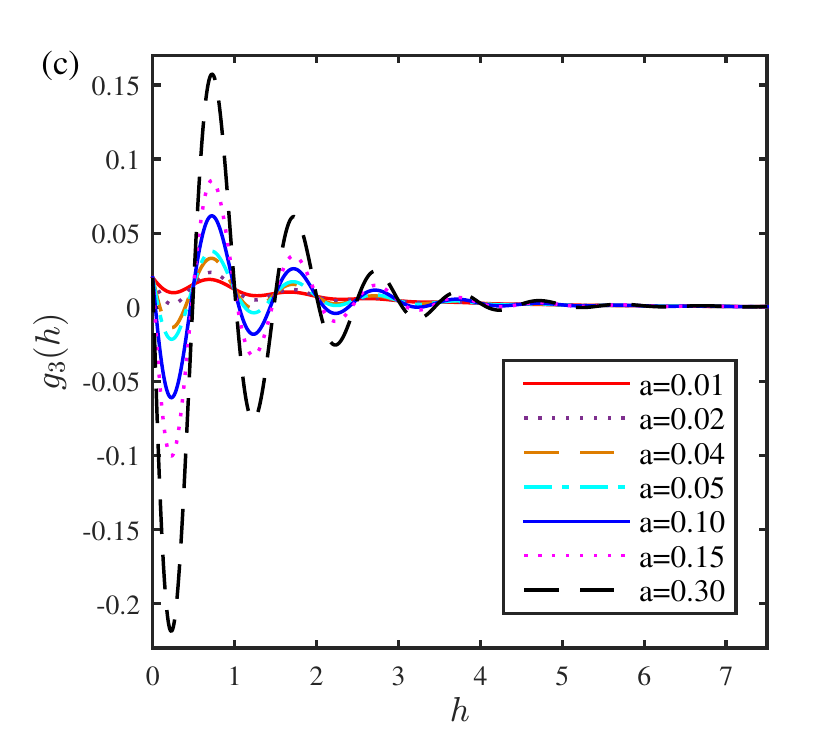}
\caption{\label{fig:moreg3} (a) A sequence of equilibrium droplet profiles with binding potential $g_3$ and varying $a$, as given in the legend. We also have $d = 0.02$, and $h_b$ is set as the lowest positive minimum of the binding potential (for $a = 0.01$, $h_b = 0.2523$, for $a = 0.02$, $h_b = 0.2402$, for $a\geq0.04$, $h_b =  0.2282$). (b) A plot of the time evolution of the normalised free energy differences of the dynamics leading to the formation of the drops in (a). (c) Shows the corresponding binding potentials.}
\end{figure*}

Fig.~\ref{fig:moreg3}(a) shows a sequence of equilibrium droplet profiles using binding potential $g_3$ with fixed $d=0.02$ and various values of $a$. The time evolution of the normalised free energy differences leading to the formation of these drops is displayed in Fig.~\ref{fig:moreg3}(b) and the corresponding binding potentials are displayed in Fig.~\ref{fig:moreg3}(c). All of these drops have the same values of $C=6$, $E=10$ and $x_f=200$. Smaller values of $a$ lead to broader droplets with flatter tops and fewer steps. This is due to the amplitude of the oscillations in the binding potential being smaller and the correspondingly higher value of the lowest positive local minima. As $a$ increases, the system undergoes a wetting transition, so the most obvious effect on the droplets is the extent of spreading, which is controlled by the equilibrium contact angle which in turn is determined by the height of the lowest positive minimum in the binding potential.

Fig.~\ref{fig:moreg3}(b) shows the corresponding normalised free energy differences for spreading on these binding potentials. As before, steps in this quantity correspond to popping events or the emergence of terraces. There are overall trends as $a$ is varied, such as secondary (and higher) terracing events are harder to see in curves for large $a$. Also, in general, the spreading to equilibrium occurs more quickly for greater $a$, as would be expected from the binding potentials and equilibrium droplet profiles, given that the distance the droplet has to spread is less. However, there are exceptions to this trend such as the case with $a=0.04$, where the `popping' event takes an unusually long time to occur. Thus, it is possible for the crossing of the normalised free energy difference curves for different values of $a$, where at a particular stage of the dynamics certain evolutions are slowed by the formation of terraces (e.g.\ by being `pinned' to particular heights), whereas at the same time such an event does not occur in an otherwise slower (larger difference between initial and equilibrium contact angles) spreading situation. We believe the occurrence of slow dynamics in the system corresponds to parts of the profile having to pass over saddles in the free energy.

Comparing the normalised free energy difference curves for $a=0.1$ and $a=0.3$ in Fig.~\ref{fig:moreg3}(b), we see the curves cross and are rather different in shape, indicating that the different stages of the dynamics occur on different timescales. Interestingly, however, both take the same overall time to finally equilibrate. The $a=0.3$ case initially decreases much more rapidly, due to the high barrier in the binding potential between the minimum at $h_0$ and the next minimum at $h_1$, corresponding to 1 layer of particles [see Fig.~\ref{fig:moreg3}(c)]. The free energy cost of having $h\approx\frac{1}{2}(h_0+h_1)$ is high, so the system chooses $h=h_0$ or $h=h_1$ as quickly as it can. Following this, there is a slower relaxation over a longer timescale. In contrast, for $a=0.1$ the energy difference between the first positive local maximum and the neighbouring minimum is much less, so there is only one timescale visible in the relaxation to equilibrium.

Having investigated the dynamics of spreading for $g_3$, a binding potential with oscillations but also having monotonic exponential decay at larger $h$, we now focus on $g_4$, which has oscillatory decay for $h\to\infty$. 

\begin{figure}[t]
\centering
\includegraphics{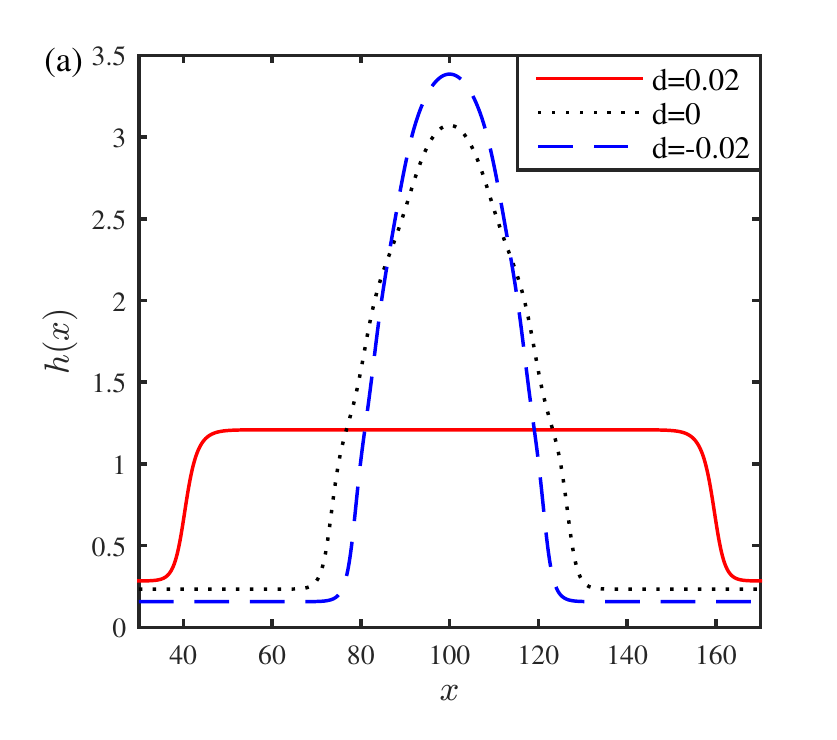}

\includegraphics{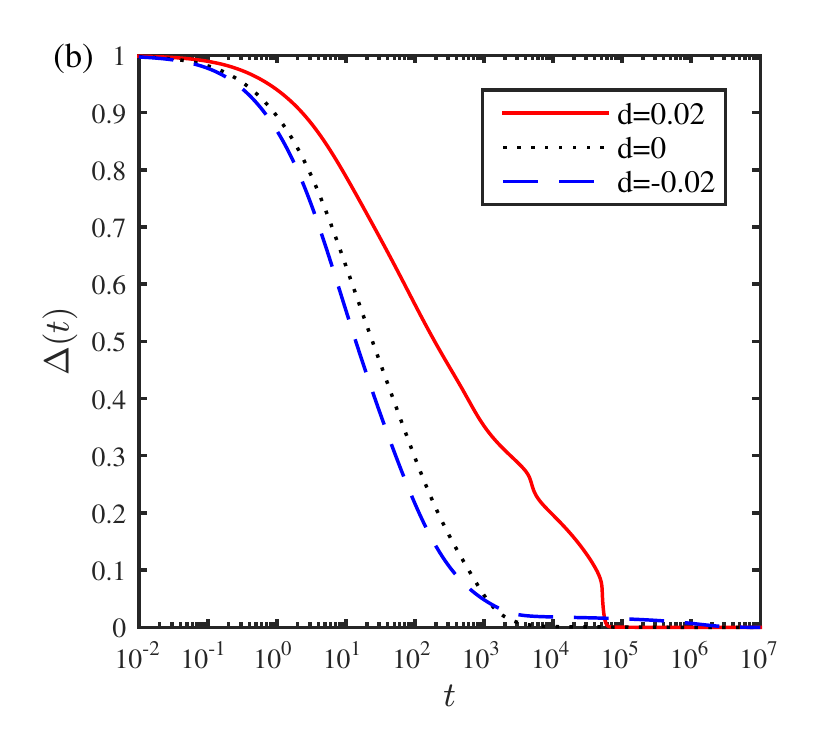}
\caption{\label{g4deq} (a) Equilibrium drop profiles with the binding potential $g_4$ in Eq.~(\ref{g4}), as the parameter $d$ is varied. We take $a = 0.01$, $b = \frac{\pi}{2}$, $c = 1$, $k = 2 \pi$ and $d = \{0.02,  0, -0.02\}$. (b) The corresponding normalised free energy difference for the spreading of the droplets.}
\end{figure}

Final equilibrium drop profiles with the binding potential $g_4$ for various values of the parameter $d$ are shown in Fig.~\ref{g4deq}(a). The initial drop profile \eqref{initialcondition} has $C=6$, $E=10$, $x_f=200$, and $h_b=\{0.3, 0.2402, 0.2282\}$ which are the corresponding minima ($h_0$) for $d=\{0.02, 0, -0.02\}$. Similar to the drop profiles with binding potential $g_3$, higher values of $d$ lead to broader droplets and flatter tops, as the system passes through the wetting transition. However, instead of having two complete layers of particles for $d=0.02$ as shown in Fig.~\ref{compareg3} for $g_3$, the final equilibrium shape with binding potential $g_4$ has only one particle layer. This is in agreement with intuitive analysis of the plot of $g_4$ against $h$ in Fig.~\ref{bp4}, since the global minimum of $g_4$ with $d=0.02$ is at $h\approx1$ and thus one layer of particles is preferred.

To investigate the dynamics of spreading with binding potential $g_4$, in Fig.~\ref{g4deq}(b) we plot the corresponding normalised free energy difference over time for the same three values of $d$. In the two $d\neq0$ cases the drop take a much longer time to reach equilibrium than the cases in Fig.~\ref{fig:moreg3} (with $g_3$). Also, there are two obvious `popping' events on the $d=0.02$ curve. At the first inflection point (at $t\approx10^3$), the drop starts to form terraces and the centre of the drop `pins' to a particular height until the first jump (`popping event') occurs between $t=10^3$ and $t=10^4$, where the normalised energy difference drops down from $\approx0.26$ to $\approx0.22$, during which the top part of the droplet flattens. A similar process occurs again at $t\approx 7\times10^4$ because one layer of particles is slightly more favourable than two or more layers of particles (see Fig.~\ref{bp4}). We believe that in this case the dynamics is particularly slow because the difference in the free energy of the different minima of $g_4$ for $d=0.02$ are rather small.

The slow dynamics for the case $d=-0.02$ in Fig.~\ref{g4deq}(b) is for a different reason. Looking at the time evolution of the drop profile (not displayed), the drop initially seems to reach equilibrium with a background film height equal to $h_0$. However, this leaves the top of the droplet on a maximum in the binding potential, and so the background film height raises up slightly to allow the top of the drop to move off the maximum. In the final equilibrium neither the background film nor the top of droplet are in any binding potential minima, but nonetheless the state is the best overall equilibrium for the entire droplet.

\begin{figure*}[t]
\centering
\includegraphics[width=0.68\columnwidth]{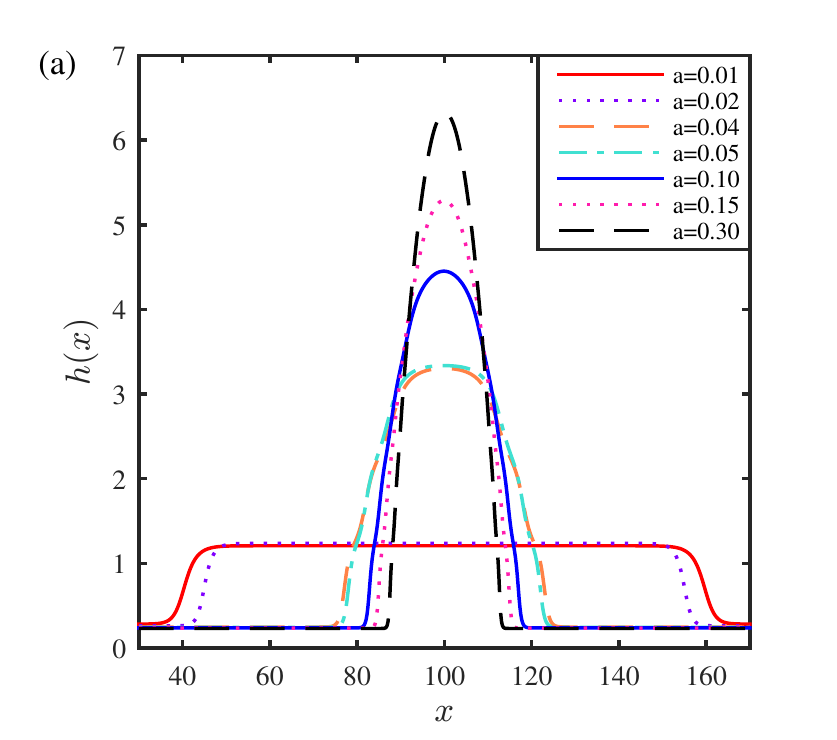}
\includegraphics[width=0.68\columnwidth]{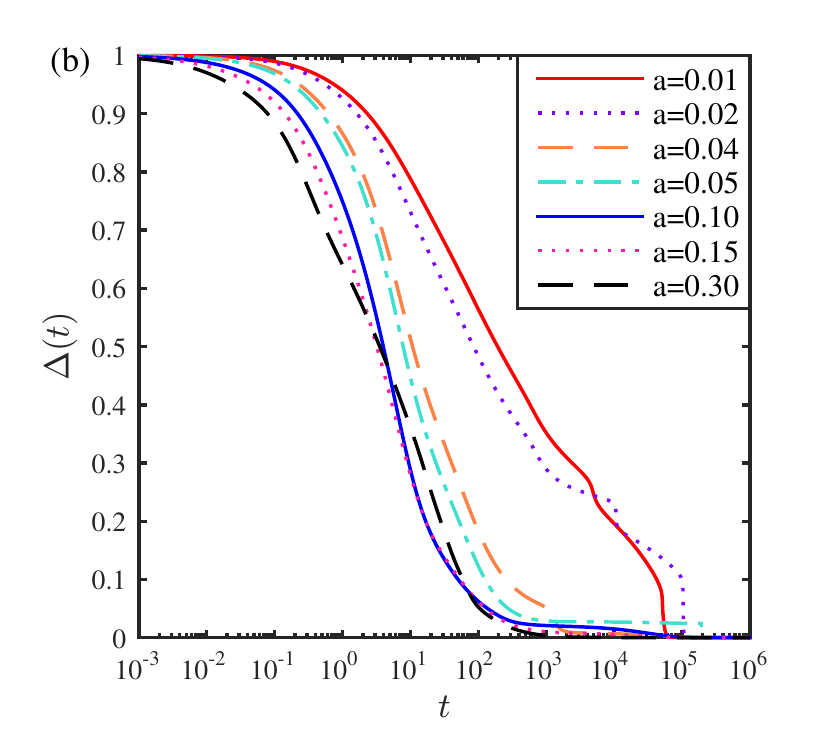}
\includegraphics[width=0.68\columnwidth]{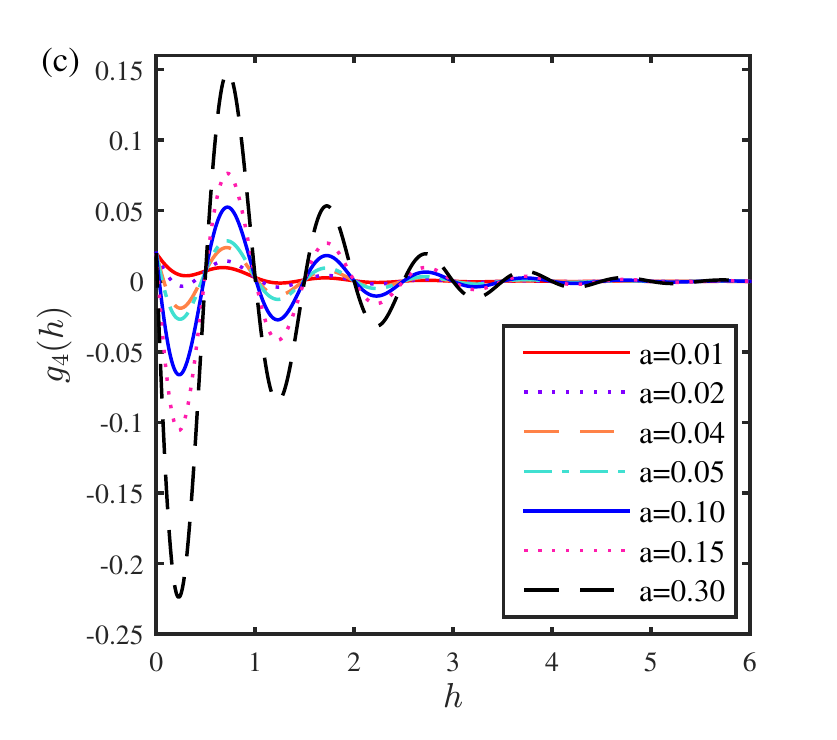}
\caption{\label{g4varya} (a) A sequence of equilibrium droplet profiles with binding potential $g_4$ and varying $a$, as given in the legend. We also have $d = 0.02$, and $h_b$ is set as the lowest positive minimum of the binding potential (for $a = 0.01$, $h_b = 0.300$, for $a = 0.02$, $h_b = 0.2642$, for $a = \{0.04, 0.05\}$, $h_b = 0.2402$, and for the remaining values of $a$, $h_b =  0.2282$). (b) A plot of the time evolution of the normalised free energy differences of the dynamics leading to the formation of the drops in (a). (c) Shows the corresponding binding potentials. }
\end{figure*}

In Fig.~\ref{g4varya}(a) we display a sequence of equilibrium droplet profiles for binding potential $g_4$ and various values of $a$. As $a$ is increased the amplitude of the oscillations in $g_4$ increases and also the the system is further from the wetting transition with a larger contact angle, since the primary minimum in $g_4$ at $h=h_0$ becomes lower as $a$ is increased. In Fig.~\ref{g4varya}(b) we display the corresponding time evolution of the normalised free energy difference and in (c) the binding potential. The overall behaviour is somewhat similar to that displayed in Fig.~\ref{fig:moreg3} for binding potential $g_3$. However, the overall time it takes to equilibrate varies even more in this case, being anywhere in the range $10^3$ -- $10^6$. As before, this is due to the slow dynamics that occurs due to popping, pinning and other such events as the droplet evolves in a complex free energy landscape having many long `valleys' and saddle points.

\subsection{Spreading versus dewetting towards equilibrium}

In view of the apparent complexity in the underlying free energy landscape in which the spreading droplet evolves, a natural question to arise is: does that landscape exhibit multiple minima? All the results presented so far correspond to spreading droplets, so to address this question, we also consider cases where the initial condition consists of a pancake-like drop that is spread out more than the expected final equilibrium state, so that the evolution towards equilibrium consists of a dewetting dynamics, with the contact line of the droplet receding.

\begin{figure}[t]
\centering
\includegraphics{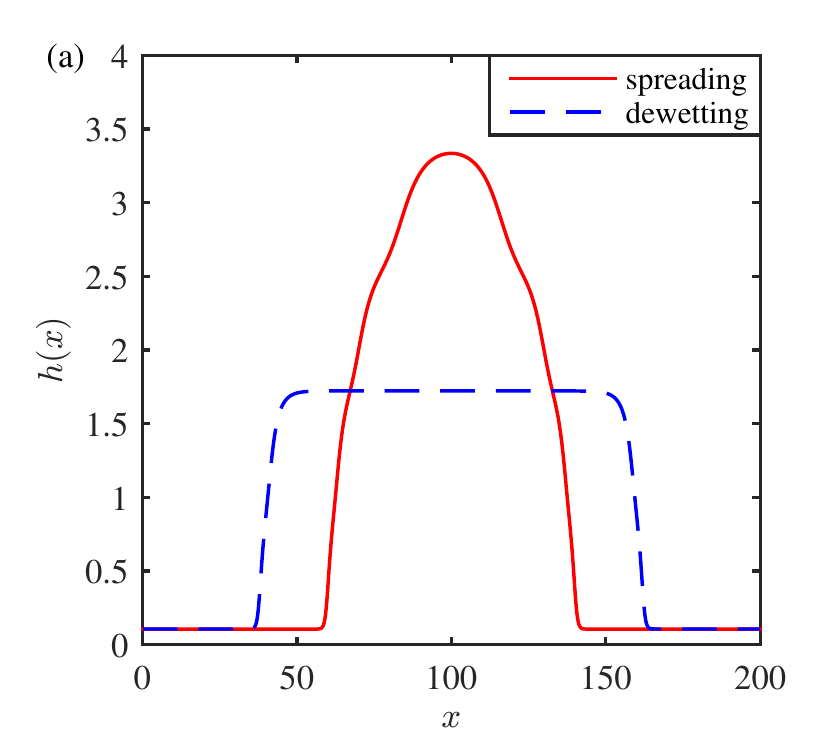}
\includegraphics{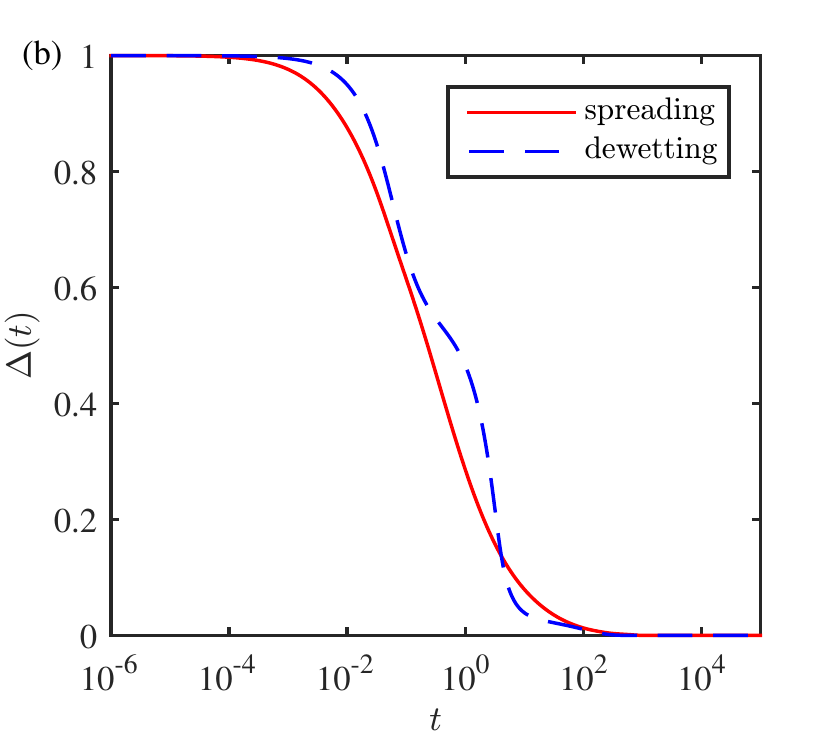}
\caption{\label{dewetting} (a) The final equilibrium state obtained for the same volume of liquid on the surface undergoing both spreading and dewetting on a background film $h=h_0$. The binding potential used is $g_2$, with initial conditions that were evolved to reach these final equilibria given in (\ref{eq:icdewet1}) for the spreading situation, and (\ref{eq:icdewet2}) for dewetting. (b) shows the corresponding evolutions of the normalised free energy differences, noting that for spreading $F(0)=16.89$ and $F(\infty)=0.046$, whereas for dewetting $F(0)=0.44$ and $F(\infty)=0.15$.}
\end{figure}

In some cases, identical equilibrium profiles are found from both spreading and dewetting simulations. However, it is also not uncommon for different equilibria to be realised. In Fig.~\ref{dewetting} we highlight a case of this latter situation, where the initial profiles for spreading and dewetting were given as
\begin{equation}
h(x, t = 0) = 18.00 \ e^{-\left[\frac{(x-x_f/2)}{6}\right]^2} + h_b,\label{eq:icdewet1}
\end{equation}
and 
\begin{equation}
h(x, t = 0) = 1.694 \ e^{-\left[\frac{(x-x_f/2)}{60}\right]^8} + h_b,
\label{eq:icdewet2}
\end{equation}
respectively, and where $x_f=200$, $h_b=h_0=0.1081$, and the binding potential $g_2$ is used. Specifically, in Fig.~\ref{dewetting}(a) we present the final equilibrium states of two droplets with the same volume that have evolved to attain different equilibrium profiles---although as expected given both simulations are for identical substrates with the same binding potentials, the effective contact angle made with the background film is seen to be in agreement in both equilibria. We further note that the drops have dynamically evolved to find the locally lowest energy configuration for their height profile across the entire domain. As the effective domain is finite (due to the periodic boundary conditions), this means that the background height at equilibrium is not exactly $h_0$, the lowest (positive) minimum value for the imposed binding potential (in this case $g_2(h)$). Indeed, the two cases in Fig.~\ref{dewetting}(a) have slightly different values for the background film height, corresponding to different values of $\lambda$ [defined in Eq.~(\ref{ga})]. In Fig.~\ref{dewetting}(b) the corresponding evolutions of the normalised free energy differences are depicted to highlight the very different approaches to equilibrium for these two situations. We further note that they have not approached the same free energy equilibrium. In particular the spreading case has been able to find a lower minimum in the energy landscape, since it has reached $F(\infty)=0.046$, compared to $F(\infty)=0.15$ for dewetting.

In our simulations, it is noticed that dewetting usually proceeds less rapidly than spreading, as has been reported previously for droplet motion simulations \cite{Savva09}. Alongside the fact that the relative depth of the minima in the binding potentials are mostly greater for smaller values of film height, we see that popping events are much less common in dewetting than in spreading and the terraces are formed in a more gradual evolution.

\subsection{Including diffusion}

All results presented thus far are for spreading onto a substrate already covered with a film of thickness $h_0$, essentially like conducting a spreading experiment on an ostensibly dry substrate but that already has a few particles adsorbed on it. We have modelled this situation assuming that the droplet evolution proceeds with advection only -- i.e.\ the case where $\bar{\alpha}_0=0$ in our governing equation (\ref{eq:ge_use}). However, as discussed in Sec.\ \ref{sec:IV}, when the amount adsorbed on the substrate is a single monolayer or less, we expect the dynamics to be diffusive. This is even more true when thin films advance onto a substrate that is completely dry, with no particles at all present on the substrate before the droplet is introduced. Therefore, we now consider the case $\bar{\alpha}_0>0$.

\begin{figure}[t]
\centering
\includegraphics{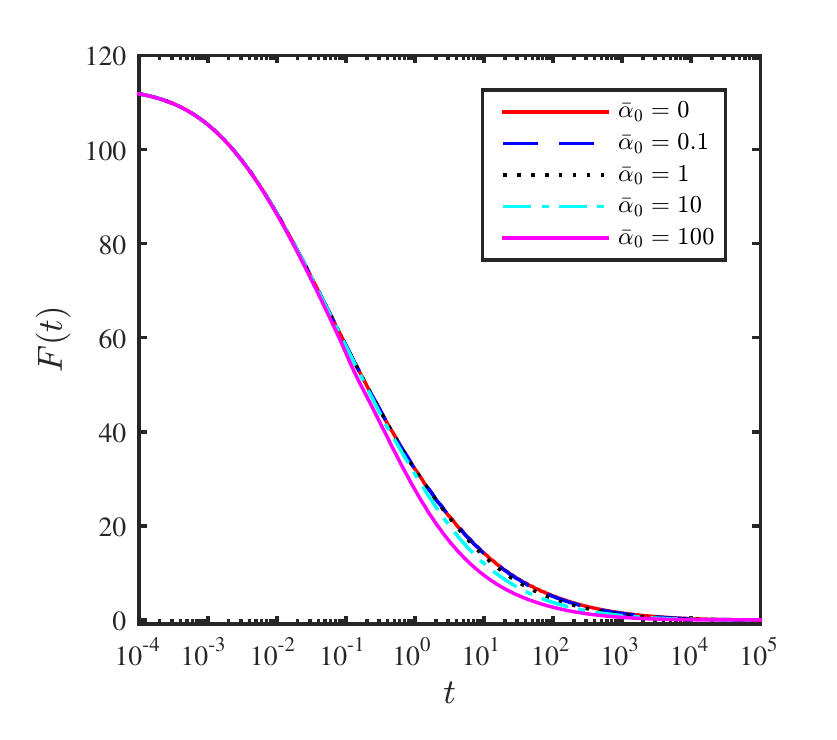}
\caption{\label{g2hb} The free energy over time during droplet spreading on a substrate with binding potential $g_2$ covered by a film of thickness $h_b\neq0$, for various values of the diffusion coefficient $\bar\alpha_0$. The initial condition has $h_b = h_0= 0.1081$, $C=60$, $E=10$ and $x_f=400$.}
\end{figure}

\begin{figure}[t]
\centering
\includegraphics{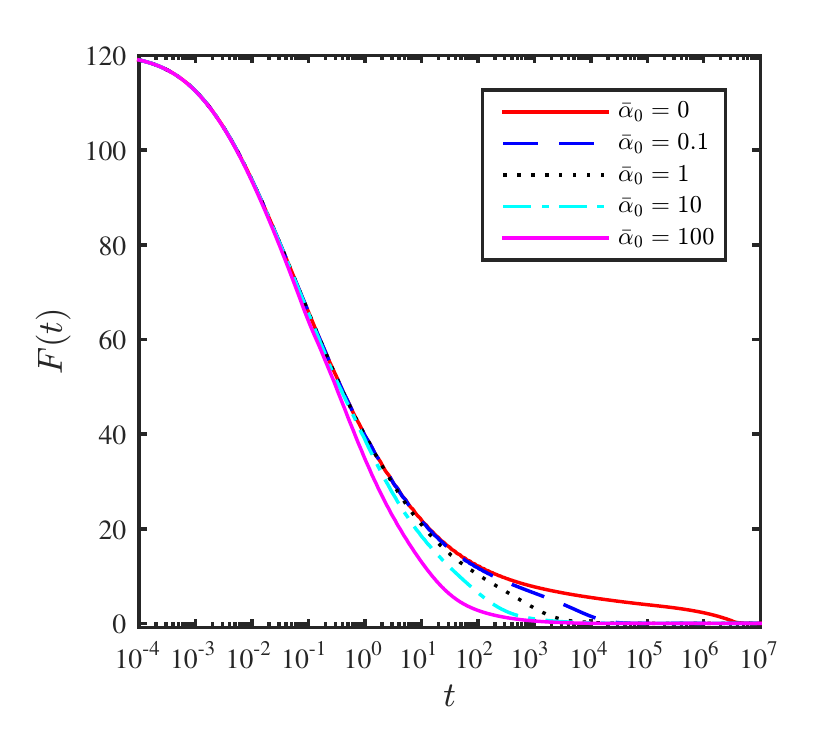}
\caption{\label{g2h0} The free energy over time during droplet spreading on a totally dry substrate with binding potential $g_2$, for various values of the diffusion coefficient $\bar\alpha_0$. The initial condition has $h_b = 0$, $C=60$, $E=10$ and $x_f=400$.}
\end{figure}

We first consider the case where the binding potential is $g_2$. In Figs.~\ref{g2hb} and \ref{g2h0} we show cases where the droplet is initiated on a background film of thickness $h_b=h_0$ (as previously) and $h_b=0$ (a totally dry initial substrate), respectively. We see two main effects: (i) increasing diffusion (larger $\bar{\alpha}_0$) speeds up the evolution in all cases; (ii) diffusion has a far greater impact when the droplet is spreading on a totally dry substrate, as in Fig.~\ref{g2h0}, compared to when spreading on an already present `precursor'. In these plots, a relatively large initial droplet is chosen (with $C=60$), hence most of the droplet has $h\gg\sigma$ throughout the evolution. Thus, as anticipated in the model development discussion in Sec.~\ref{sec:IV}, diffusion does not have a dramatic effect for large drops initially and $O(1)$ times. However, for drops of any size, the latter stages of the approach to equilibrium ultimately requires a reshaping of the height profile along terraces (at small multiples of $\sigma$) caused by the oscillatory binding potential. In these latter stages, diffusion then speeds up this reshaping process, and even for large droplets can decrease the time to reach equilibrium by orders of magnitude (e.g.\ see the behaviour of the free energy in Fig.~\ref{g2h0} for $F(t)\lesssim 5$).

\begin{figure}[t]
\centering
\includegraphics{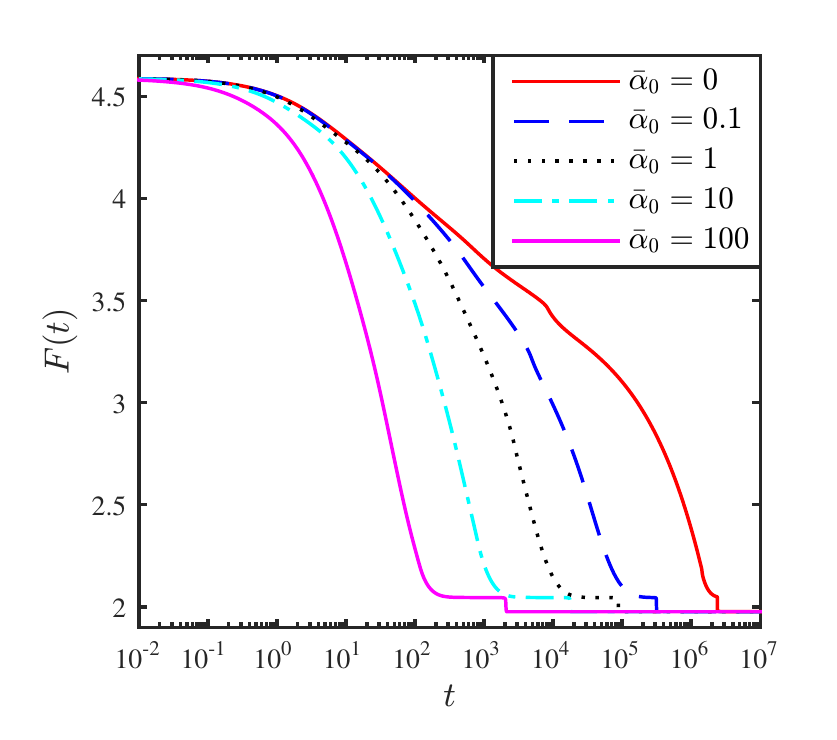}
\caption{\label{g3h0} The free energy over time during droplet spreading on a totally dry substrate with binding potential $g_3$, for various values of the diffusion coefficient $\bar\alpha_0$. The initial condition is $h_b = 0$, $C=6$, $E=10$ and $x_f=200$.}
\end{figure}

\begin{figure}[t]
\centering
\includegraphics{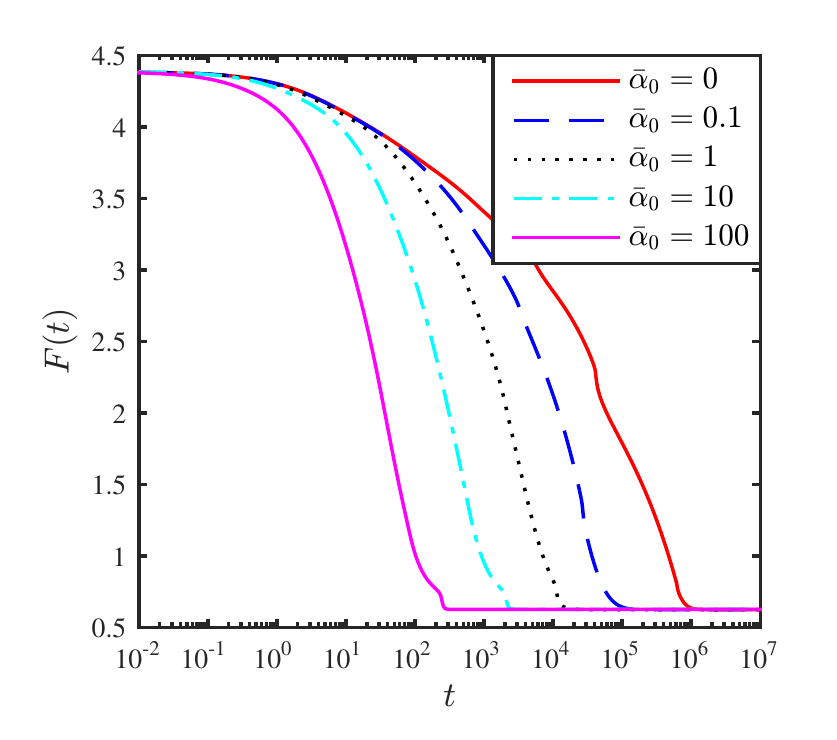}
\caption{\label{g4h0} The free energy over time during droplet spreading on a totally dry substrate with binding potential $g_4$, for various values of the diffusion coefficient $\bar\alpha_0$. The initial condition is $h_b = 0$, $C=6$, $E=10$ and $x_f=200$.}
\end{figure}

For much smaller droplets, however, where the average height of the drop is small, the diffusion is much more influential across the entire evolution, speeding up the equilibration. In Figs.~\ref{g3h0} and \ref{g4h0} we plot the evolution of the free energy for a particular set of parameters in $g_3$ and $g_4$ for a small initial droplet on a totally dry substrate ($C=6, h_b=0$). In these cases the spreading happens more rapidly when $\bar\alpha_0>0$. It can be many orders of magnitude faster than the case $\bar\alpha_0=0$. This shows that (i) including diffusion is essential in situations where the physics dictates that it has an effect, and that (ii) the order in which parts of the droplets evolve can be reversed. For example, consider the extreme cases of $\bar{\alpha}_0=0$ and $\bar{\alpha}_0=100$ in Fig.~\ref{g4h0}. For high diffusion the early time dynamics consists of the precursor foot spreading out rapidly, before a final popping event to reach equilibrium for the larger part of the droplet centre. In contrast, when there is no diffusion $\bar{\alpha}_0=0$, the spreading occurs as usual with the main part of the droplet evolving to very close to the final shape before the precursor foot eventually is formed in the final approach to equilibrium.

\section{Concluding remarks}
\label{sec:VI}

In this paper we have investigated thin liquid films spreading on a flat solid substrate, including effects such as surface tension, oscillatory binding potentials, advective flow dynamics and surface diffusion. Lubrication theory and dimensional analysis have been used to derive a model governing equation \eqref{everything} for the drop height profile. From this one can also obtain the pressure and velocity profile. Solving numerically using the Finite Difference Method has allowed us to simulate the thin film droplet spreading.

The oscillatory binding potentials that we have used model the molecular packing that can occur in certain liquids at interfaces \cite{de2013capillarity, brader2003statistical, heslot1989molecular, YaKB1992pra, hughes2015determining, hughes2016liquid, DeCh1974jcis,KrDe1995cpl}. These occur in systems that exhibit layering transitions near to the wetting transition. Note that spreading nanoparticle-laden drops \cite{WaNi2003n} have previously been modelled with thin-film models incorporating oscillatory disjoining pressures \cite{CNWT2004jcis,matar2007dynamic, hu2012influences}. These dynamical models describe the time evolution of two coupled fields: the height of the liquid film and the local concentration of nanoparticles. In Ref.~\cite{matar2007dynamic} it is shown that the presence of nanoparticles can also lead to terraced droplets and steps emerging from the contact line. In the context of the gradient dynamics form presented here in \eqref{diffusioneq}  it should be noted that such a form also exists for nanoparticle-laden or surfactant-laden films \cite{ThAP2012pf,ThTL2013prl,ThAP2016arxiv}. Such a model shows that a oscillatory nanoparticle-dependent wetting potential does not only result in an oscillatory disjoining pressure but also in an correspondingly amended chemical potential for the particles.

We have shown that having an oscillatory binding potential leads to a rich and varied droplet spreading dynamics. The time evolution towards equilibrium can often exhibit several stages. There is the usual spreading and relaxation from the initial condition, but there is also the formation of terraces and `popping' events where there is a rapid drop in the free energy due to a jump of the droplet free surface from one minimum of the binding potential to another. There is also the usual final long-time approach to equilibrium. We believe this rich behaviour is due to the complexity of the underlying free energy landscape that exhibits multiple minima, long valleys along which the dynamics is slow and saddle points. To better understand the underlying free energy landscape, we expect a systematic phase plane analysis is required -- i.e.\ a systematic examination how diagrams like that in Fig.~\ref{stream} vary as the parameters in the system are changed. Work in this direction is currently under way.

Our extended thin-film hydrodynamic model \eqref{everything} is also capable of describing the crossover from advective to diffusive dynamics that must occur when the film thickness is of order one particle thick or less. Such a crossover must occur \cite{ala2002collective}. We have shown that when diffusion is included, the droplet spreading is faster, particularly for very small droplets and for all droplets in the latter stages of their approach to equilibrium. Incorporating diffusion can also change the dynamic pathway taken -- c.f.\ the discussion above around Fig.~\ref{g4h0} and Ref.~\cite{honisch2015instabilities} where the nonlinear dynamics of the Plateau-Rayleigh instability of a liquid ridge is investigated comparing pathways occuring for different mobility functions. Note however that incorporating a small amount of diffusion does not change Tanner's law \cite{tanner1979spreading}. Recall that Tanner's law states that over a significant portion of the spreading time of a radially symmetric liquid drop (puddle), the radius $R$ grows in time as $R\sim t^n$, with exponent $n=1/10$. For 2D spreading drops like those studied here, the law still applies, but with exponent $n=1/7$ \cite{tanner1979spreading}. We have checked that including moderate diffusion does not change Tanner's law, however it does change the pre-factor, so that the overall time for the drop to equilibrate is less with diffusion incorporated. The extent of the time period that Tanner's law holds (usually after initial transient relaxations to a quasistatic shape up until a change to exponential behaviour during the latter stages of approach to equilibrium) can also be significantly reduced in our oscillatory binding potential simulations. This happens, as would be expected, when the majority of the droplet profile lies in the region where terraces form and the greater range of dynamical features occur, creating the rich evolutions we have explored. A final note on the relevance of Tanner's law, however, is that if a very large diffusion coefficient is imposed then the overriding asymptotic structure of the bulk of the droplet moving with a $h^3$ mobility would break down, and in this situation we could expect an entirely different evolution. We leave this diffusion dominated regime for possible future work.

The work presented here has largely focused on the simplified oscillatory binding potentials $g_3(h)$ and $g_4(h)$. Recall that the more complex $g_2(h)$ in Eq.~\eqref{bp2} is the one that was obtained as a fit to the DFT data \cite{hughes2015determining}. Additionally, all of these have the Hamaker constant $H=0$. There is therefore clearly much more work to do understanding the spreading behaviour when realistic binding potentials are used that are valid for all values of film thickness. Comparison of different binding potentials could also be performed for a wider range of fluids, e.g. colloidal fluids, oils, polymeric solutions etc.

Other extensions to the present work that would be fruitful include solving for the droplet spreading dynamics in 3D, including gravity, e.g.\ to also consider sliding droplets, which exhibit extremely rich behaviour \cite{engelnkemper2016morphological}. Preliminary results indicate that it would also be interesting to study what happens when the initial condition is not symmetric, and also the droplet dynamics in the presence of other droplets.

\appendix*

\section{Convergence test}

As a prerequisite to generating the results presented above, we of course conducted convergence tests in order to be certain of the accuracy of the solutions of our numerical analysis. There are some interesting results from this analysis relating to the choice of spatial grid spacing in the discretisation that readers should be aware of if trying to reproduce our results.

\begin{figure}[t]
\centering
\includegraphics{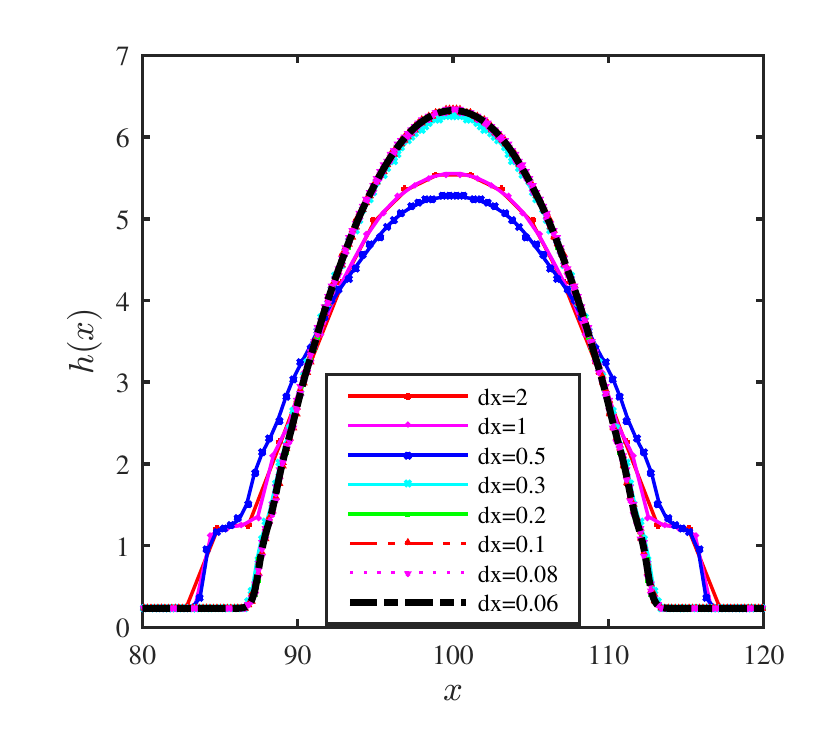}
\caption{\label{a03}  Equilibrium droplet profile with $g_3$ and with $x_f=200$, $d = 0.02$, $a = 0.3$, $h_b = 0.2282$, discretising with various grid spacings $dx$ to obtain the numerical solution, where $dx = \frac{x_f}{N}$ and where $N$ is the total number of grid points. {The values used are $N=100$, $N=200$, $N = 400$, $N = 667$, $N = 1000$, $N = 2000$, $N=2500$, and $N=3333$. The last four of these are virtually indistinguishable.}}
\end{figure}

To test accuracy, we calculate droplet evolutions for a series of different mesh discretisations, going to increasingly finer meshes (i.e.\ an increase of the number of points within the interval) and compare the results with the previous one. If the results are equal within a small percentage error, the first mesh is good enough. On the other hand, if the results differ by a large amount, the same process must be repeated for a finer mesh. A finer mesh generally results in a more accurate solution and lowers the convergence error, however it requires progressively larger memory and takes more time to compute -- particularly for the time evolution given the effective number of ODEs to be solved increases with the number of grid points.
Thus a desirable mesh would combine acceptable accuracy with economical cost.

Fig.~\ref{a03} shows the sequence of the droplet profiles for $g_3$ with $d=0.02$, $a = 0.3$, {$C=6$ and $E=10$}, with different $dx$, where $dx$ is the grid spacing $dx = \frac{x_f}{N}$, where $N$ is the number of grid points and is chosen as {$N=100$, $N=200$, $N = 400$, $N = 667$, $N = 1000$, $N = 2000$, $N=2500$, and $N=3333$. There is a greatly elongated terrace of height $h(x)\approx1$ for the curves with the coarsest three discretisations, which disappears as $dx$ decreases. Also, the terraces in these curves are more pronounced compared to the others. Thus in this case, a poor mesh grid results in a very different final equilibrium droplet shape. There is also sometimes} a loss of volume during the time evolution when the grid spacing is too large, i.e.\ the algorithm does not accurately capture the conservation of mass, which must be satisfied in our non-volatile system. 

%\begin{acknowledgments}
\section*{Acknowledgements}
We are grateful to Svetlana Gurevich for the discussions about phase plots like Fig.~\ref{stream}.
%\end{acknowledgments}

%\bibliography{bibliography}


\end{document}